\documentclass[twocolumn,showpacs,preprintnumbers,amsmath,amssymb,
groupedaddress,prc,floatfix]{revtex4}
\usepackage{mathrsfs}
\usepackage{longtable}
\usepackage{graphicx}
\usepackage{graphics}
\usepackage{epsfig}
\usepackage{dcolumn}
\usepackage{rotating}

\begin{document}

\title{Single Particle Configurations of the Excited States of $^{203}$Po}

\author{S. Chatterjee}
\affiliation{UGC-DAE Consortium for Scientific Research, Kolkata Centre, Kolkata 700098, India}
\author{A. Ghosh}
\altaffiliation{Present Address: ONGC, Cinnamara Complex, Jorhat, Assam 785704, India}
\affiliation{UGC-DAE Consortium for Scientific Research, Kolkata Centre, Kolkata 700098, India}
\author{D. Arora}
\altaffiliation{Present Address: Inter University Accelerator Centre, New Delhi 110067, India}
\affiliation{UGC-DAE Consortium for Scientific Research, Kolkata Centre, Kolkata 700098, India}
\author{S. Das}
\affiliation{UGC-DAE Consortium for Scientific Research, Kolkata Centre, Kolkata 700098, India}
\author{B. Mondal}
\affiliation{UGC-DAE Consortium for Scientific Research, Kolkata Centre, Kolkata 700098, India}
\author{S. Samanta}
\affiliation{UGC-DAE Consortium for Scientific Research, Kolkata Centre, Kolkata 700098, India}
\author{R. Raut}
\altaffiliation{Corresponding Author: rajarshi.raut@gmail.com}
\affiliation{UGC-DAE Consortium for Scientific Research, Kolkata Centre, Kolkata 700098, India}
\author{ S. S. Ghugre}
\affiliation{UGC-DAE Consortium for Scientific Research, Kolkata Centre, Kolkata 700098, India}
\author{P. C. Srivastava}
\affiliation{Department of Physics, Indian Institute of Technology, Roorkee, Roorkee 247667, India}
\author{A. K. Sinha}
\altaffiliation{Presently: Hon. Adjunct Professor, Department of Physics, Savitribai Phule Pune University, Pune, India}
\affiliation{UGC-DAE Consortium for Scientific Research, University Campus, Khandwa Road, Indore 452017, India}
\author{U. Garg}
\affiliation{Department of Physics, University of Notre Dame, Notre Dame, Indiana 46556}
\author{H. K. Singh}
\affiliation{Department of Physics, Indian Institute of Technology, Bombay, Mumbai 400076, India}
\author{Neelam}
\affiliation{Department of Physics and Astrophysics, University of Delhi, New Delhi 110007, India}
\author{K. Rojeeta Devi}
\affiliation{Department of Physics and Astrophysics, University of Delhi, New Delhi 110007, India}
\author{A. Sharma}
\affiliation{Department of Physics, Himachal Pradesh University, Shimla 171005, India}
\author{S. S. Bhattacharjee}
\affiliation{Inter University Accelerator Centre, New Delhi 110067, India}
\author{R. Garg}
\affiliation{Inter University Accelerator Centre, New Delhi 110067, India}
\author{I. Bala}
\affiliation{Inter University Accelerator Centre, New Delhi 110067, India}
\author{R. P. Singh}
\affiliation{Inter University Accelerator Centre, New Delhi 110067, India}
\author{S. Muralithar}
\affiliation{Inter University Accelerator Centre, New Delhi 110067, India}
\date{\today}

\begin{abstract}

Excited states of the $^{203}$Po ($Z = 84, N = 119$) have been investigated after populating them
through $^{194}$Pt($^{13}$C,4n) fusion-evaporation reaction at E$_{beam}$ = 74 MeV and using a 
large array of Compton suppressed HPGe clover detectors as the detection setup for the emitted
$\gamma$-rays. Standard techniques of $\gamma$-ray spectroscopy have been applied towards
establishing the level structure of the nucleus. Twenty new $\gamma$-ray transitions have been
identified therein, through $\gamma-\gamma$ coincidence measurements, and spin-parity assignments 
of several states have been determined or confirmed,
following the angular correlation and linear polarization measurements on the observed $\gamma$-rays.  
The excited states have been interpreted in the framework of large basis shell model calculations, 
while comparing their calculated and experimental energies. They have been principally ascribed to 
proton population in the $h_{9/2}$ and $i_{13/2}$ orbitals outside the $Z = 82$ closure and 
neutron occupation of the $f_{5/2}$, $p_{3/2}$ and $i_{13/2}$ orbitals in the $N = 126$ shell.

\end{abstract}

\pacs{23.20.Lv,21.10.Hw,21.60.Cs}

\maketitle

\section{Introduction}

The shell model of the nucleus has remained its most credible microscopic description through 
more than seven decades now. Testing the model across the nuclear chart and refining the inputs,
towards accomplishing better overlap with data, has been an agenda of nuclear structure 
studies through their evolving practice. The exercise is facilitated by developments in  
computational resources that help circumvent the dimensional challenges incurred in the application
of shell model, particularly to heavier systems such as those around Pb ($Z = 82$).
It may be noted that the very validity of the shell model for describing level structures 
around the proton $Z = 82$ closure
was a subject of early investigations in the region. While the closure at $Z = 82$ was 
identified to be sufficiently stable against collective excitations \cite{Rah85}, it was also observed that
light Hg ($Z = 80$) isotopes do exhibit collectivity and there were predictions of similar phenomena
in the proton-rich side of the ($Z = 82$) closure, for the light Po ($Z = 84$) nuclei \cite{Wec85}. 
The studies undertaken towards resolving the proposition, however, froze on describing the excitations
of light-Po isotopes, such as $^{199-201}$Po, within the framework of the shell model. This was also
commensurate with the systematically calculated \cite{Wec85} shapes of the Pb isotopes starting from 
$^{208}$Pb ($Z = 82, N = 126$) and extending to the lighter ones. The doubly-magic $^{208}$Pb, quite 
expectedly, exhibited deep energy minimum for a spherical shape; the minimum became shallower for 
lighter systems in the isotopic chain and eventually evolved into a double minima corresponding to
both prolate and oblate deformations for nuclei as light as $^{190}$Pb ($Z = 82, N = 108$). 
Such a scenario, however, wasn't established in $^{198}$Pb or $^{202}$Pb that still manifested
near spherical shapes and it was found valid to interpret the excitation schemes of the neighboring
light Po isotopes from the perspectives of the shell model. The merits of such interpretation 
notwithstanding, it was largely extracted from the evolution of experimentally 
observed level energies and their spacings across the isotopic and/or the isotonic chains. That was 
presumably owing to the limited wherewithal then available for computational endeavors but, nevertheless,
could provide insights into the particle excitations underlying the level scheme of the nuclei being 
studied. The experimental findings in these studies mostly followed population of the nuclei of interest
in $\alpha$- or heavy-ion induced fusion-evaporation reactions and detection of the $\gamma$-rays using
modest setups of few Ge detectors and, at times, using conversion electron measurements alongwith. \\

The only existing precedence of spectroscopic study of the $^{203}$Po ($Z = 84, N = 119$) nucleus, 
following its population in a fusion-evaporation reaction, was by Fant {\it{et al.}} \cite{Fan86}.
The nucleus was populated using $\alpha$-induced reaction on $^{204}$Pb and the de-excitation $\gamma$-rays
were detected using small planar Ge(Li) detectors, large coaxial Ge(Li) detectors and intrinsic Ge detectors.
Conversion electrons were also measured in conjunction. The level scheme of the nucleus was established 
upto an excitation energy of $\sim$ 4.4 MeV and spin $\sim$ 18$\hbar$. However, only a selected number of 
$\gamma$-ray transitions, presumably the strongest ones, and levels were identified above the 25/2$^+$ state; 
the spin-parity assignments were considerably tentative therein. The configurations of the excited states 
were largely ascribed to the coupling of an odd neutron hole to the excitations of the even $^{204}$Po-core ($Z = 84, N = 120$).
Two configurations, based on proton excitations outside the closed proton shell of the $^{208}$Pb-core, were
identified in the latter. These were $\pi h_{9/2}^2$ and $\pi h_{9/2}i_{13/2}$ that resulted in maximum spins
8 and 11 respectively. The available single particle orbitals for the odd neutron are $2f_{7/2}, 1h_{9/2}, 1i_{13/2}, 
3p_{3/2}, 2f_{5/2}, 3p_{1/2}$ and the first $5/2^-, 3/2^-, 1/2^-, 13/2^+$ states, in $^{203}$Po, were identified with
single neutron excitations therein. The 17/2$^+$, 21/2$^+$, and 25/2$^+$ yrast states in odd-A Po isotopes
were attributed to the odd neutron hole $\nu i_{13/2}^{-1}$ coupled to the excitations of the corresponding Pb-core 
or of the two valence protons of the Po-core, resulting in states 2$^+$ - 8$^+$. This followed the systematics
of the yrast states in odd-A Pb and Po isotones. It may be noted that the yrast 17/2$^+$ and the 21/2$^+$ states
in isotopes $^{199-205}$Pb had been ascribed to pure neutron excitations, such as $\nu p_{1/2}^{-1}f_{5/2}^{-1}i_{13/2}^{-1}$
and $\nu f_{5/2}^{-2}i_{13/2}^{-1}$. However, such (pure neutron) excitations would result into states of
higher excitation energies than those of the yrast 17/2$^+$ and the 21/2$^+$ levels in odd-A Po isotopes. It was thus
found reasonable to assign the pure neutron excitations to the respective non-yrast states. The 27/2$^+$ and the 29/2$^+$
levels in the odd Po nuclei were identified with three-quasiparticle configurations $\pi h_{9/2}^2\otimes\nu i_{13/2}^{-1}$.
The configurations for the isomeric 25/2$^-$ and 29/2$^-$ were derived from their overlap with the systematics
of these states observed in the Pb isotopes. Accordingly, their configuration in $^{203}$Po was identified
to be similar to that in $^{201}$Pb and the same is $(\pi (h_{9/2}^2)_{0^+}\otimes\nu p_{1/2}^{-2}f_{5/2}^{-3}(i_{13/2}^{-2})_{12^+})_{25/2^{-}29/2^{-}}$. The findings in $^{203}$Po thus upheld the interpretation of 
its excitation scheme within the framework 
of the single particle excitations, as had been established for the still lighter isotopes 
of the nucleus \cite{Wec85}. This was also a 
continuing trend from the heavier isotopes such as $^{205,207}$Po \cite{Rah85}. The absence of collectivity was 
further corroborated by the absence of enhanced B(E2) in these nuclei \cite{Fan86}. \\

The present paper reports a spectroscopic investigation of the level structure of $^{203}$Po, using 
updated experimental facilities as well as contemporary framework for shell model calculations. The objective
was to explore possible features in the excitation scheme of the nucleus, through the use of a
large array of high-resolution gamma-ray detectors in the setup, and to test the reproducibility of the observed
level energies in the calculations of the interacting shell model. The computational exercise is a validation
of the model Hamiltonian used for the purpose as well as of facility in identifying and quantifying the single
particle excitations that contribute to the observed level scheme. \\

\section{Experimental Details and Data Analysis}

Excitations of the $^{203}$Po nucleus were investigated following its population in the $^{194}$Pt($^{13}$C,4n) reaction
at $E_{lab}$ = 74 MeV. The target was 13 mg/cm$^2$ thick self-supporting foil of enriched (99\%) $^{194}$Pt. The 
beam was delivered by the 15 UD Pelletron at IUAC, New Delhi and the beam energy was so chosen after an excitation
function measurement at the commencement of the experiment. As per the predictions of the statistical model calculations,
at this beam energy, the aforementioned reaction would be of dominant cross section amongst the possible 
compound nucleus fusion-evaporation channels while the fission (exit) channel would amount to 
$\sim$ 25\% of the total fusion cross-section. Indeed, the yield of $^{203}$Po was observed to be maximum 
when compared to the other fusion-evaporation products, that principally included isotopes of Po ($Z = 84$), 
Bi ($Z = 83$) and Pb ($Z = 82$), as illustrated in Fig. 1. 
The detection system was the Indian National Gamma Array (INGA) setup
at IUAC \cite{Mur10} 
and (then) consisted of eighteen Compton suppressed HPGe clover detectors positioned at 148$^o$ (4 detectors),
123$^o$ (4 detectors), 90$^o$ (6 detectors), 57$^o$ (2 detectors), and 32$^o$ (2 detectors). An assembly
of three absorber sheets of lead, tin, and copper was afixed on the face of the heavymet collimator of the 
Anti Compton Shield (ACS) in each detector. The absorbers facilitated in reducing the intensity of the X-rays,
from the thick target, being incident on the detectors (and thus contributing in the event trigger). 
Data was principally acquired under the condition that at least two Compton suppressed HPGe clover 
detectors needed to fire in coincidence for generating the event trigger. The two- and higher-fold events
acquired was $\sim$ 2$\times$10$^9$.\\

\begin{figure}
\includegraphics[angle=-90,scale=.35,trim=2.0cm 2.0cm 0.0cm 1.0cm,clip=true]{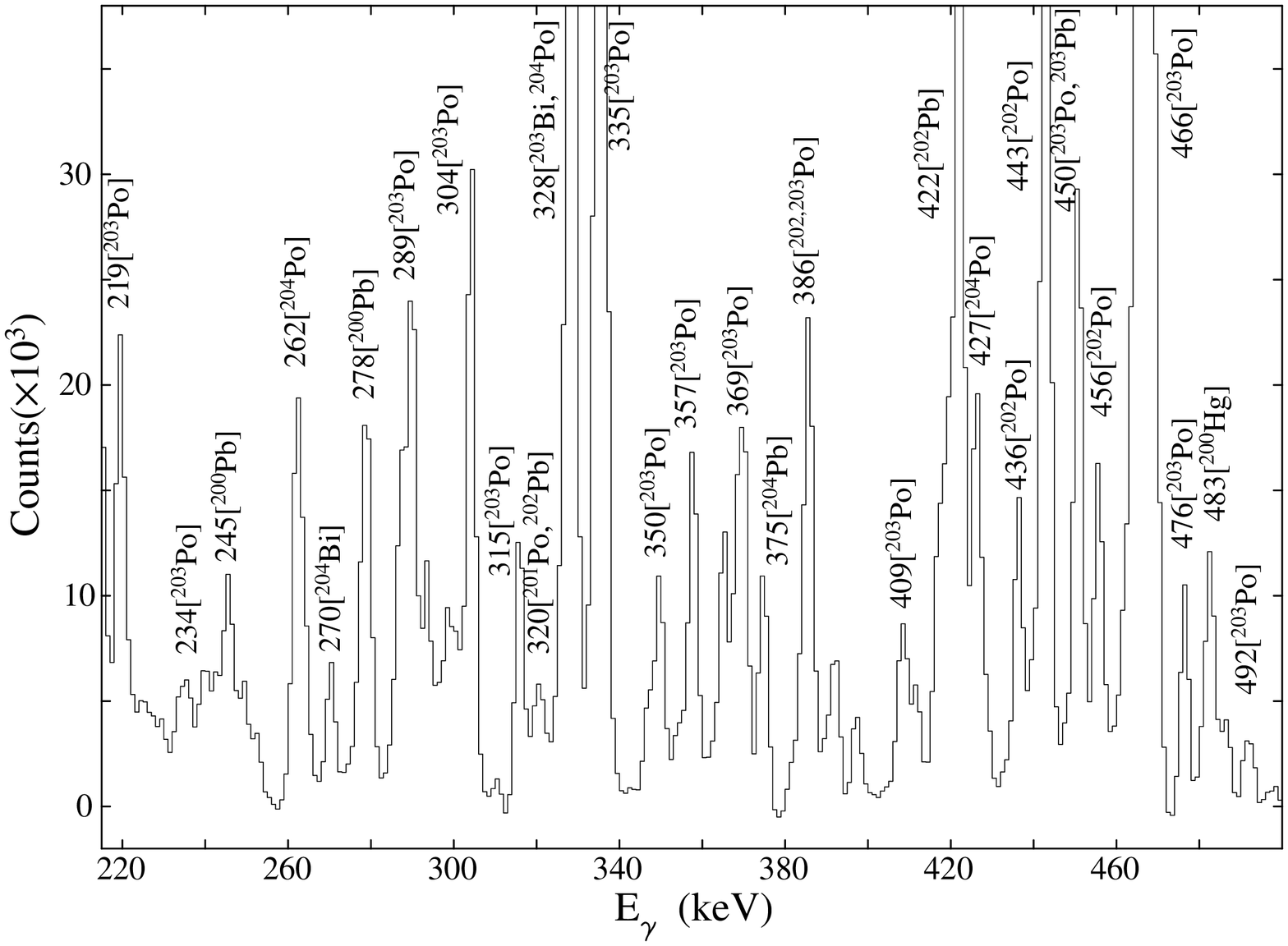}
\caption{\label{fig1} Part of the $\gamma$-ray spectrum corresponding to the full projection of a 
$\gamma$-$\gamma$ symmetric matrix and illustrating the different product nuclei populated 
in the present experiment.}
\end{figure}

The data was sorted into spectra, symmetric and asymmetric (angle dependent) $\gamma$-$\gamma$ matrices as 
well as $\gamma$-$\gamma$-$\gamma$ cube using SPRINGZ \cite{Das17_2} and INGASORT \cite{Bho01} codes 
and subsequently analyzed using the RADWARE \cite{Rad95} package. The methodology and the 
objectives of the exercise were identical to that of any regular investigation of nuclear level structure
using $\gamma$-ray spectroscopy. These have been detailed in numerous papers, such as Ref. \cite{Sam18,Sam19},
and are briefly mentioned herein. The coincidence relationships between the observed $\gamma$-ray transitions 
were extracted from the symmetric $\gamma$-$\gamma$ matrix and $\gamma$-$\gamma$-$\gamma$ cube. The
coincidences along with the intensity considerations were applied for the placement of the $\gamma$-ray
transitions in the level scheme of the nucleus. The assignment of multipolarities of the
$\gamma$-rays followed determination of their $R_{ADO}$ (Ratio of Angular Distribution from Oriented Nuclei)
values using

\begin{equation}
R_{ADO} = \frac{I_{\gamma1} \ at \ 32^o \ (Gated \ by \ \gamma_2 \ at \ all \ angles)}{I_{\gamma1} \ at \ 123^o \ (Gated \ by \ \gamma_2 \ at \ all \ angles)}
\end{equation}

\noindent{where $I$ is the intensity of the transition (of interest, $\gamma_1$ in the above equation) 
in the relevant gated spectrum that is generated
from the appropriate angle dependent matrix. As far as this analysis is concerned, the $R_{ADO}$ 
value for the stretched dipole ($\Delta$J = 1) transitions is 0.73$\pm$0.01 while for the 
stretched quadrupole ($\Delta$J = 2) ones, 
it is 1.34$\pm$0.01. These values were derived from $R_{ADO}$s of transitions with previously 
established multipolarities and belonging to other Po isotopes populated in the same experiment. 
The $R_{ADO}$ values determined for different $\gamma$-ray transitions, observed in this study, 
are represented in Fig. 2.} \\

\begin{figure}
\includegraphics[angle=-90,scale=.35,trim=1.0cm 1.0cm 0.0cm 1.0cm,clip=true]{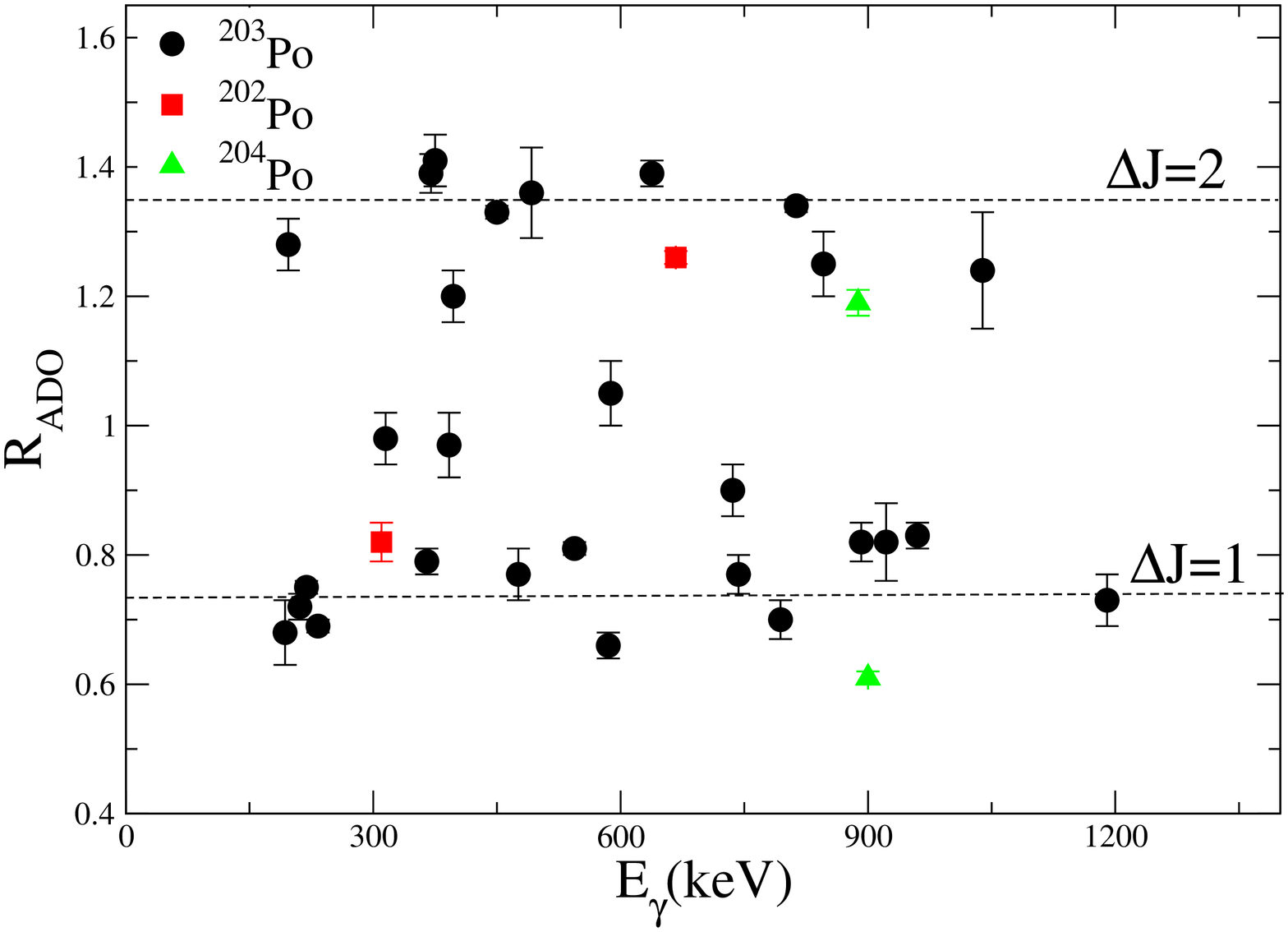}
\caption{\label{fig2}$R_{ADO}$ values for transitions of $^{203}$Po, as determined 
in the current analysis. Those for selected transitions of $^{202,204}$Po are plotted 
as reference.}
\end{figure}

\begin{figure}
\includegraphics[angle=-90,scale=.30,trim=1.0cm 1.0cm 0.0cm 1.0cm,clip=true]{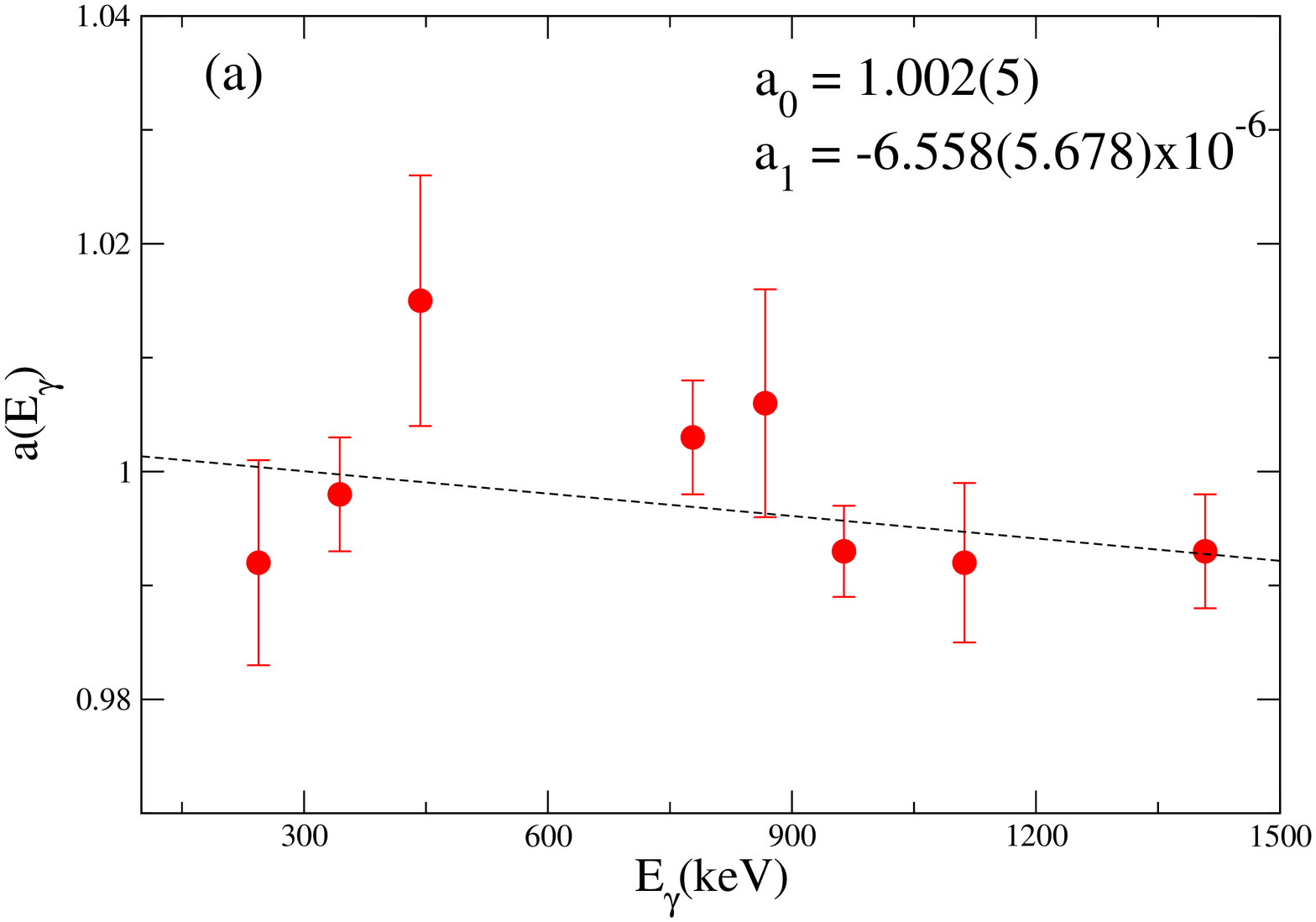}
\includegraphics[angle=-90,scale=.30,trim=1.0cm 1.0cm 0.0cm 1.0cm,clip=true]{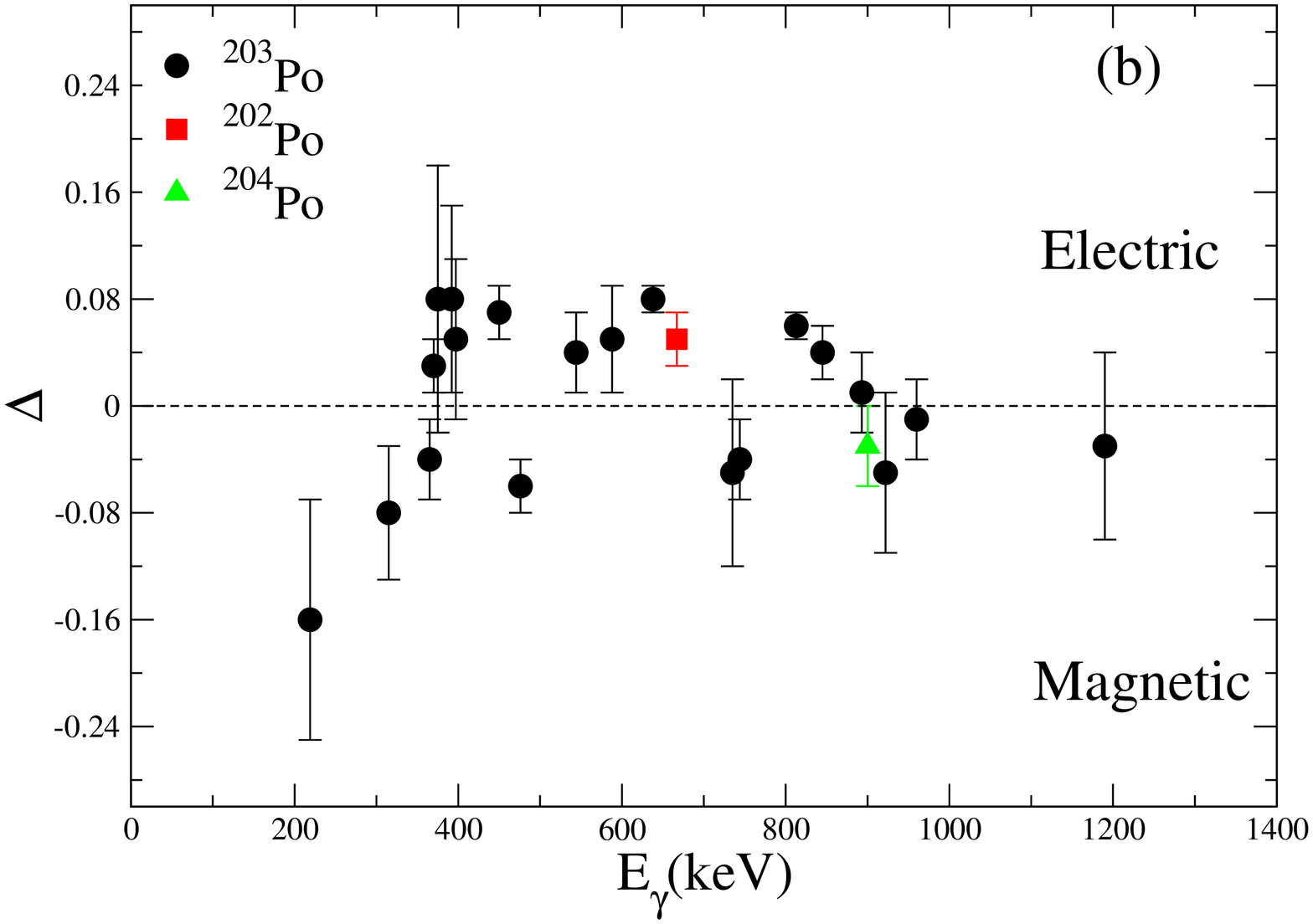}
\includegraphics[angle=-90,scale=.30,trim=1.0cm 1.0cm 0.0cm 1.0cm,clip=true]{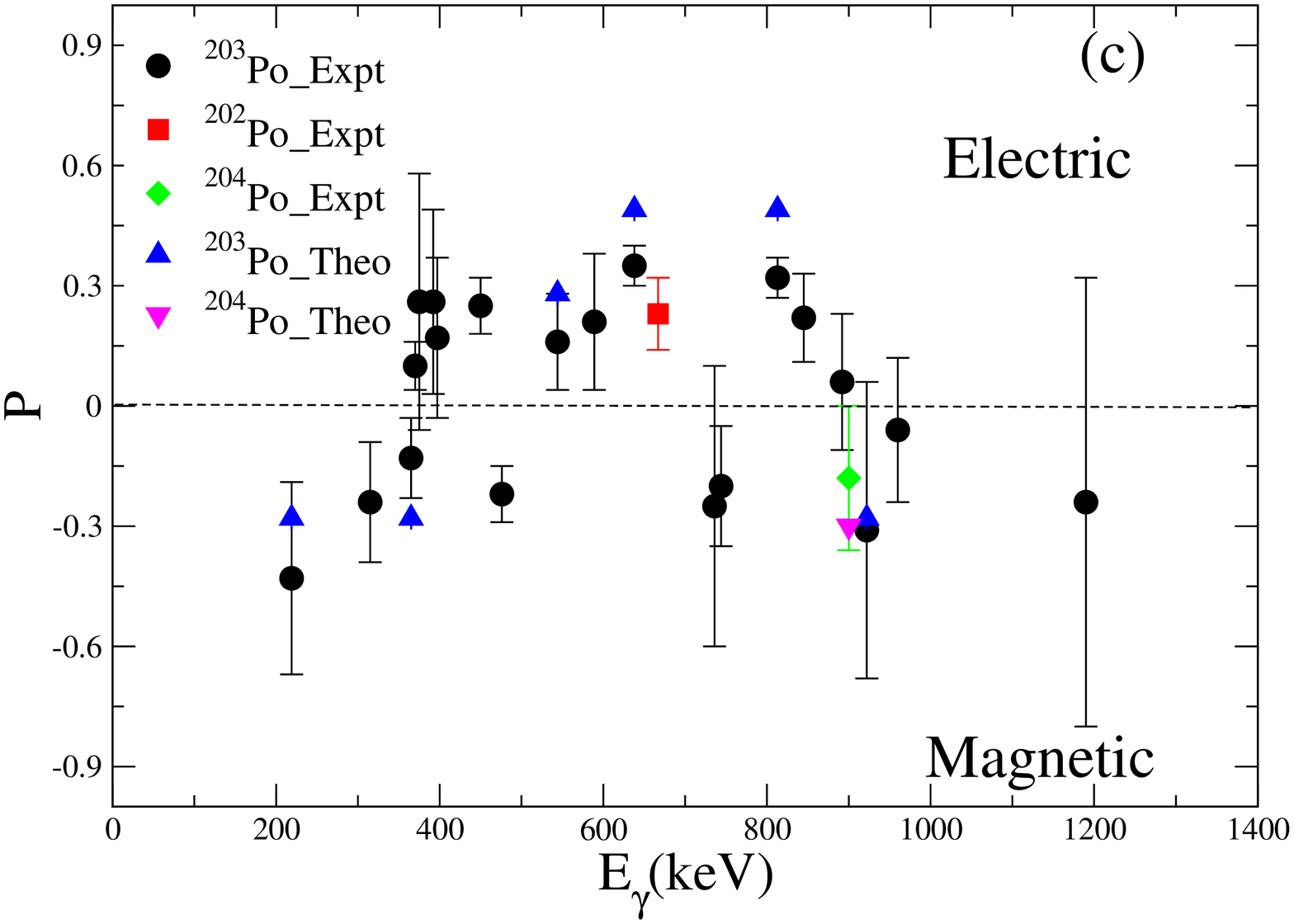}
\caption{\label{fig3}(a) Plot of geometrical asymmetry as a function of $\gamma$-ray energy.
(b) Polarization asymmetry of transitions of $^{203}$Po. (c) Linear 
polarization values for transitions of $^{203}$Po along with the corresponding theoretical estimates for
some of them (of pure multipolarity). The $\Delta$ and $P$ values for selected transitions of other isotopes, that were 
populated in the same experiment, are included for validation.}
\end{figure}

The electromagnetic nature of the transitions were assigned on the basis of their polarization asymmetry
evaluated using,

\begin{equation}
\Delta = \frac{aN_\perp \ - \ N_\parallel}{aN_\perp \ + \ N_\parallel}
\end{equation}

\noindent{where $N_\perp$ and $N_\parallel$ are respectively the number of photons of the $\gamma$-ray of interest
that are scattered perpendicular to and parallel to the reference plane. The latter is defined by 
the beam direction and the direction of emission of the $\gamma$-ray. Each of the four crystals of a HPGe 
clover detector operates as scatterer while the two adjacent ones, parallel and perpendicular to the
scatterer, operate as absorbers and facilitate in identifying the scattering events in the respective 
directions. The asymmetry between the two scattering possibilities is known to be maximum at 90$^o$. 
Thus, the $N_\perp$ ($N_\parallel$) for $\gamma$-rays is extracted from a
matrix that has been constructed with the perpendicular (parallel) scattering events in the detectors 
at 90$^o$ on one axis and the coincident detections in detectors at all other angles on the
other axis. The coincidences aid in the unambiguous identification of the $\gamma$-ray transition
being analyzed. The $a$ in Eq. (2) represents the asymmetry that is characteristic to the geometry 
of the detection setup. It was determined from the asymmetry between $N_\perp$ and $N_\parallel$ 
for $\gamma$-rays of (unpolarized) radioactive sources, such as $^{152}$Eu, and using $a = N_\parallel/N_\perp$.
The typical plot of $a$, as a function of $\gamma$-ray energy, for the present setup is 
illustrated in Fig. 3(a). The observed asymmetry between $N_\perp$ and $N_\parallel$ for polarized
$\gamma$-rays, such as those emitted by spin oriented ensemble of nuclei produced in 
fusion-evaporation reactions, depends on the degree of their polarization ($P$) and the sensitivity ($Q$)
of the measurement setup. These are related through,}

\begin{equation}
P = \frac{\Delta}{Q}
\end{equation}

\noindent{with,}

\begin{equation}
Q(E_\gamma) = Q_0(E_\gamma)(CE_\gamma \ + \ D)
\end{equation}

\noindent{where,}

\begin{equation}
Q_0(E_\gamma) = \frac{\alpha + 1}{\alpha^2 + \alpha + 1}
\end{equation}

\noindent{$\alpha$ being $E_\gamma/m_ec^2$, $m_ec^2$ is the electron rest mass energy. The $C$ and $D$ 
parameters for the purpose were adopted from those following the work by Palit {\it{et al.}} \cite{Pal00} 
and are $C = 0.000099 \ keV^{-1}$ and $D = 0.446$.} \\

As per the regular methodology of nuclear structure studies, using $\gamma$-ray spectroscopy, 
the information on coincidence relationships between the $\gamma$-rays along with their 
intensities, multipolarities and electromagnetic nature, as resulting from the aforementioned analysis,
were used to identify the excitation scheme of the nucleus and the same is discussed in the
next section.

\section{Results}

\begin{turnpage}
\begin{figure*}
\includegraphics[angle=-90,scale=1.00,trim=4.0cm 0.0cm 6.0cm 1.0cm,clip=true]{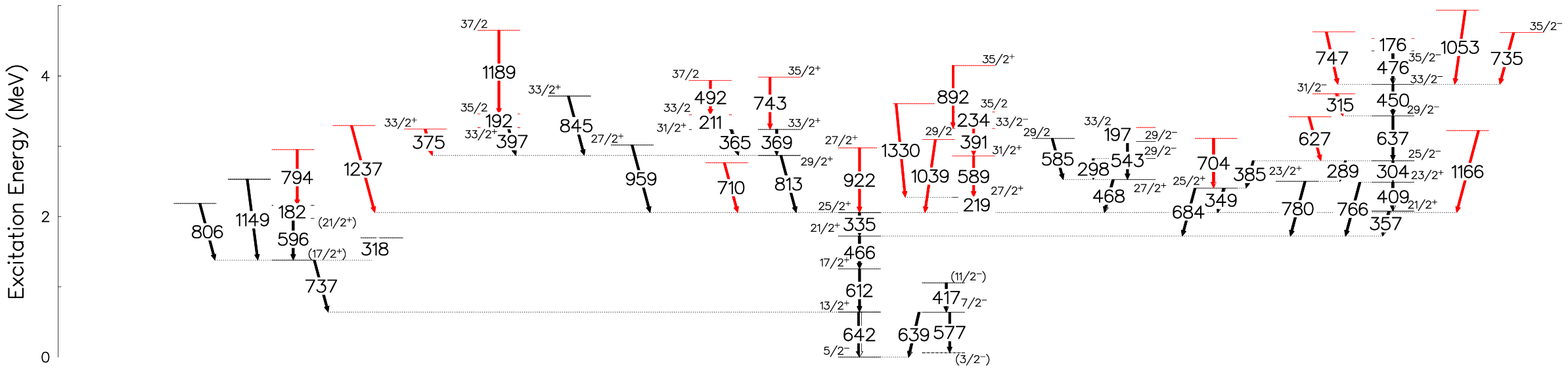}
\caption{\label{fig4} Excitation scheme of $^{203}$Po following the present work. The $\gamma$-ray 
transitions that have been newly identified in this study are indicated in red.}
\end{figure*}
\end{turnpage}
\clearpage

The excitation scheme of $^{203}$Po, as established or confirmed in the present investigation, is 
illustrated in Fig. 4. Figs. 5 and 6 illustrate the representative gated spectra respectively projected from 
$\gamma$-$\gamma$ matrix and $\gamma$-$\gamma$-$\gamma$ cube. The observed coincidences 
have been used to identify the placement of transitions in the level scheme.
Twenty new $\gamma$-ray transitions have been placed in the level scheme of the nucleus and the 
following modifications have been made in the existing \cite{Fan86,nndc} 
assignments therein. The details of the $\gamma$-ray transitions are recorded in Table I. 
(The energies of the transitions and the levels
are rounded off to the nearest integer in the discussions herein.)

\begin{figure}
\includegraphics[angle=-90,scale=.35,trim=3.0cm 1.0cm 0.0cm 0.0cm,clip=true]{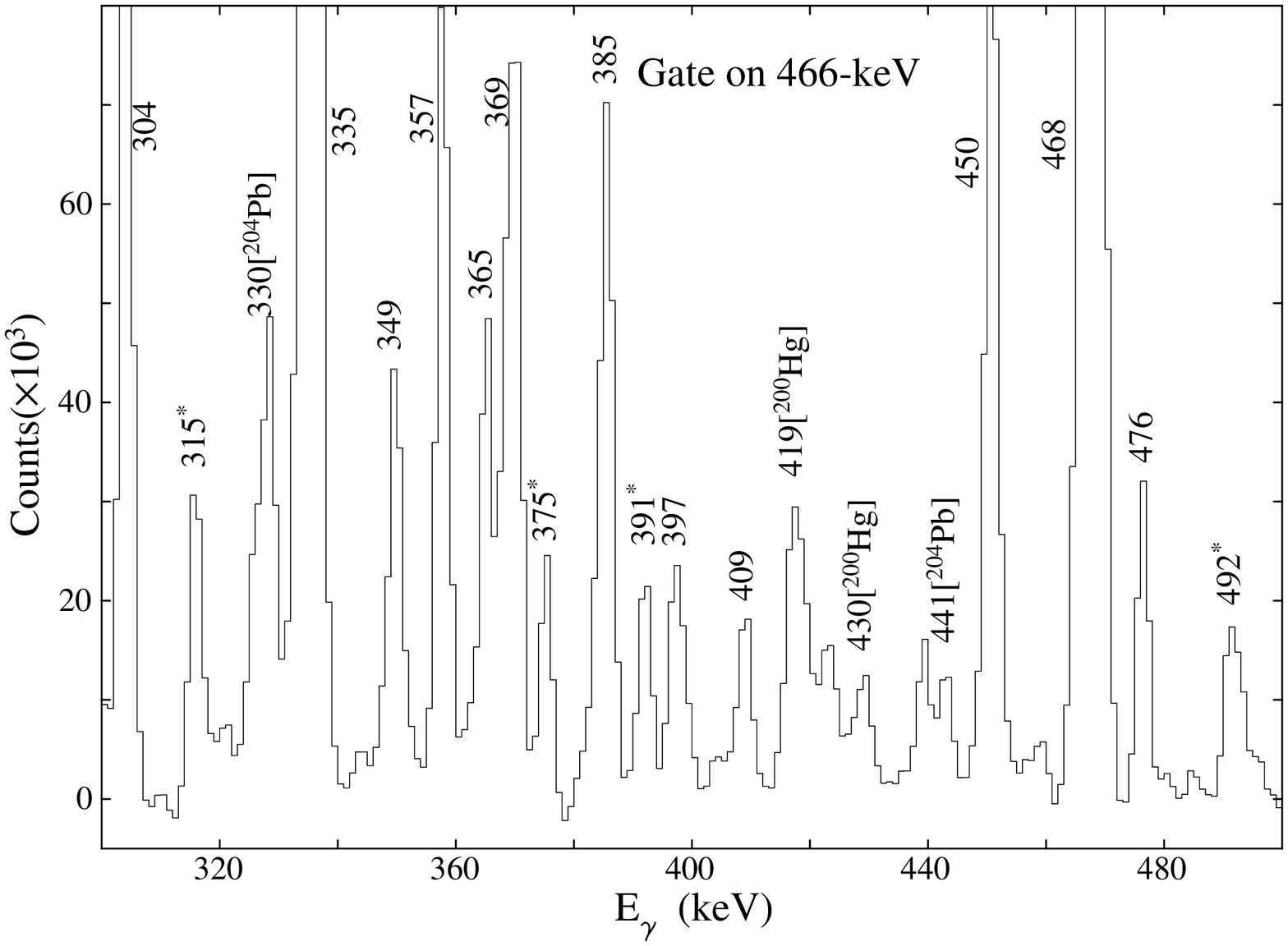}
\includegraphics[angle=-90,scale=.35,trim=3.0cm 1.0cm 0.0cm 0.0cm,clip=true]{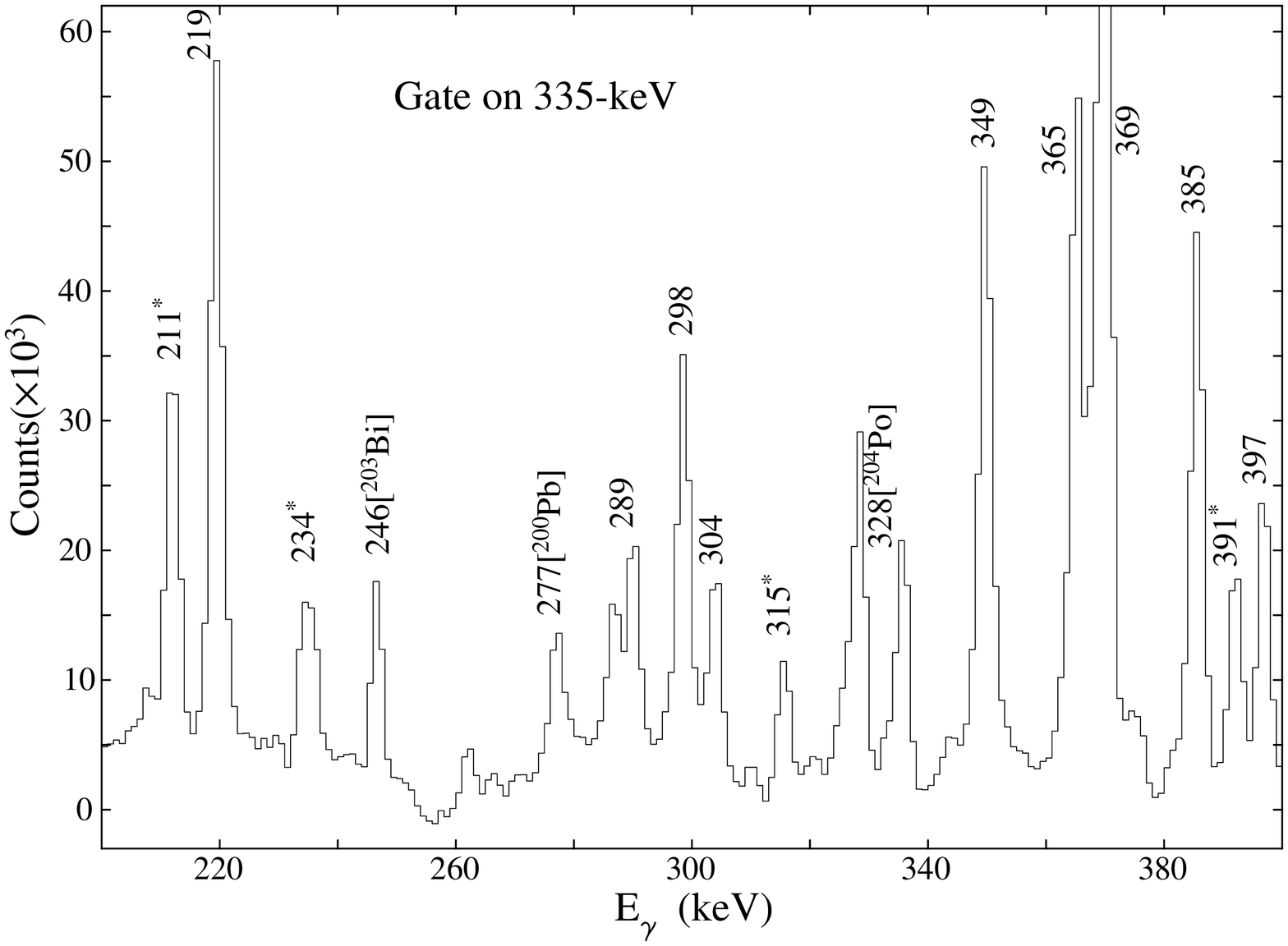}
\caption{\label{fig5}Representative spectra projected out of $\gamma$-$\gamma$ matrix with gate on transitions of $^{203}$Po, as indicated in the inset of the respective spectrum. The $\gamma$-rays newly identified in the present work are marked with *. Those resulting from overlapping coincidences in other nuclei, populated in the same experiment, are also labeled accordingly.}
\end{figure}

\begin{figure}
\includegraphics[angle=-90,scale=.35,trim=1.0cm 1.0cm 0.0cm 0.0cm,clip=true]{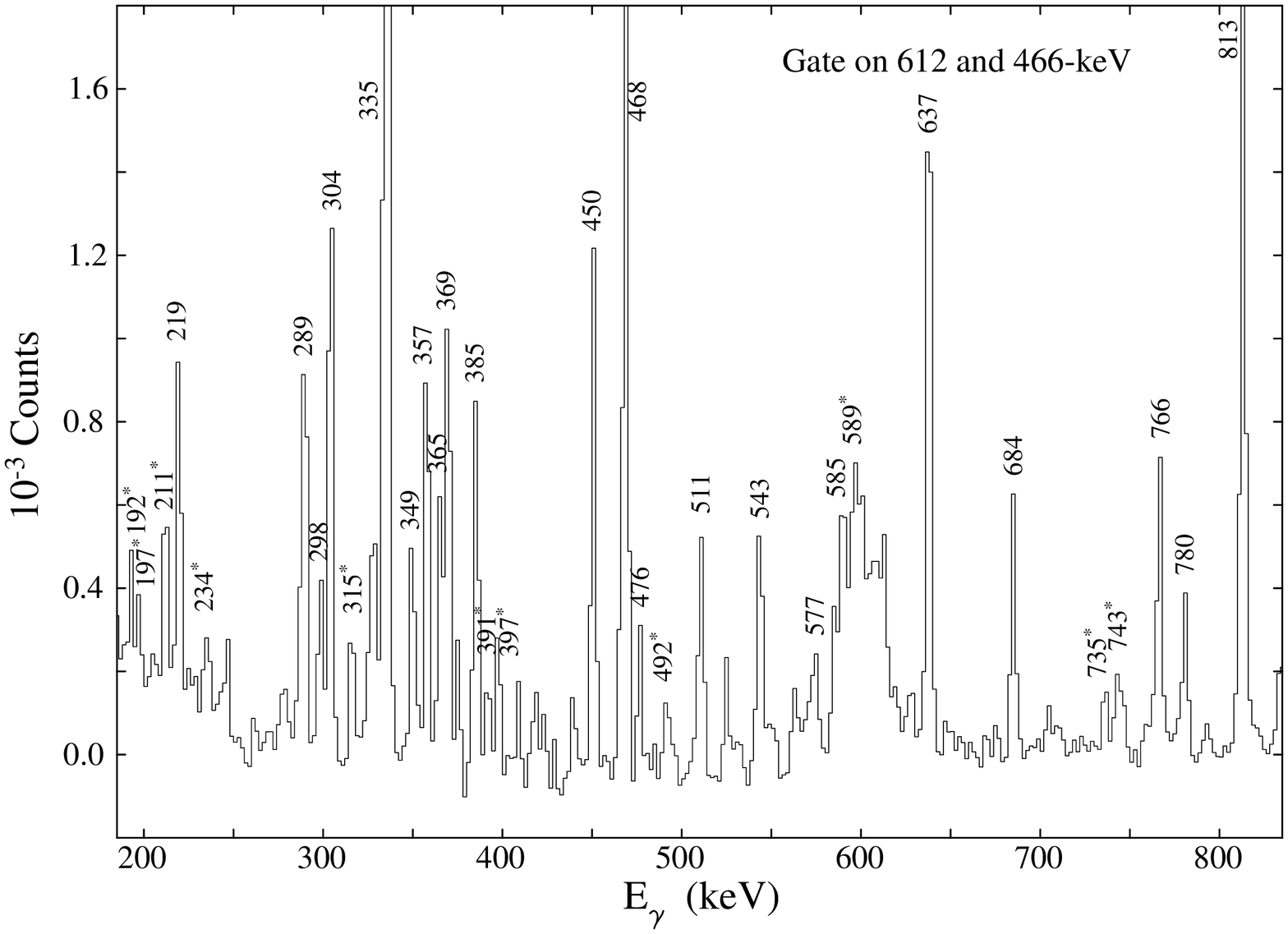}
\caption{\label{fig6}Representative spectrum projected out of $\gamma$-$\gamma$-$\gamma$ cube with double gate on transitions of $^{203}$Po, as indicated in the inset of the spectrum. The $\gamma$-rays newly identified in the present work are marked with *.}
\end{figure}

\begin{enumerate}

\item{The placement of 397-keV transition has been changed with respect to that assigned in the 
literature, as de-exciting a $\sim$ 1527-keV level \cite{Sem87}. The level and the $\gamma$-ray 
was not reported by Fant {\it{et al.}} while in the present study the placement of the transition
has been modified to one de-exciting the $\sim$ 3264-keV state. The level has been marked as a new one
in the level scheme (Fig. 4) while the $\gamma$-ray transition is identified to have been observed
previously, albeit with a different placement.}

\item{The 219-keV transition de-exciting the 2274-keV level has been identified as a M1 and the state
has been identified to be of spin-parity 27/2$^+$. There was no spin-parity assignment for this level, 
identified as $\sim$ 2277-keV by Fant {\it{et al.}} \cite{nndc}, in the previous studies.}

\item{The 959-keV transition, de-exciting the 3014-keV state, has been assigned a mixed M1+(E2) 
nature, following the present measurements. Accordingly, the state has been assigned a spin-parity of
27/2$^+$ that is at variance with the assignment by Fant {\it{et al.}} \cite{nndc}. The latter
had identified the $\gamma$-ray as a pure E2 one and had tentatively assigned the spin-parity of the 
state ($\sim$ 3018-keV, as per Fant {\it{et al.}}) as (29/2$^+$).}

\item{The 543-keV $\gamma$-ray, from the 3066-keV state, has been assigned a multipolarity of 
E1 in this study. It was tentatively identified as M1, by Fant {\it{et al.}}, and the spin of the
level (at $\sim$ 3070-keV, as per Fant {\it{et al.}}) was accordingly assigned to be 29/2.}

\item{The 585-keV transition de-exciting the 3108-keV state has been established as a pure dipole
in this study. However, the electromagnetic nature of the same could not be unambiguously determined in the 
present investigation. The multipolarity of the transition was undetermined in the work by 
Fant {\it{et al.}} and consequently there was no spin assignment for the state (at $\sim$ 3112-keV, 
as quoted by Fant {\it{et al.}}) therein.}

\item{The spin-parity of the 3236-keV state, de-excited by the 369-keV transition, has been confirmed 
to be 33/2$^+$ in this study. The assignment for the level (at $\sim$ 3241-keV, as per Fant {\it{et al.}}) 
was only tentative in the previous work \cite{nndc}.}

\item{The spin-parity of the 3712-keV level has been assigned as 33/2$^+$ in this work, following the E2 assignment
of the 845-keV $\gamma$-ray that de-excites the state. There was no multipolarity assignment for the
transition or spin-parity assignment for the state (at $\sim$ 3717-keV, as per Fant {\it{et al.}}) 
in the previous studies \cite{nndc}.}

\item{The 3877-keV state has been assigned spin-parity of 33/2$^-$, in this measurement. This is 
following the identification of the 450-keV transition, that de-excites the level, 
as an E2 one herein. Previously \cite{nndc}, the transition
was assigned as M1 and the spin-parity of the state (at $\sim$ 3882-keV, as quoted by Fant {\it{et al.}})
as 31/2$^-$. Fig. 7 represents the spectra of the transition corresponding to the perpendicular
and the parallel scattering events and illustrates the dominance of the former that leads to 
positive value of polarization asymmetry (Eq. 2) or polarization (Eq. 3).} 

\item{The spin-parity of the 4352-keV level has been confirmed to be 35/2$^-$ in the present work. The 
assignment was tentative for the state (at $\sim$ 4358-keV, quoted by Fant {\it{et al.}}) in the 
previous studies. The 476-keV transition, de-exciting the state, has been identified as M1 in this study and this is 
different from the E2 assignment by Fant {\it{et al.}}.}

\end{enumerate}

An additional proposition can be put forth on the multipolarity assignment of the 182-keV transition de-exciting
the 2156-keV state. Since the state is known to be an isomer of $T_{1/2}$ $>$ 200 ns \cite{Fan86}, the multipolarity 
of the transition could not be ascertained from its $ADO$ ratio and its polarization asymmetry. These measurements 
are valid for transitions emitted by spin oriented ensemble of nuclei, such as produced in fusion-evaporation 
reactions, while the aforementioned isomeric lifetime is sufficient to induce dealignment. 
If the observed intensity of this 182-keV transition is corrected for electron conversion, using
codes such as BrICC \cite{Kib08}, it is $\sim$ 60\% increased if the transition is an E2 one and $\sim$ 300\%
enhanced if it is a M1. The latter would result in an unbalanced intensity across the 1975-keV state that is
fed by the 182-keV transition and de-excited by the 596-keV one. If the 182-keV transition is thus interpreted
to be of E2 nature, the 2156-keV level can be assigned a spin-parity of 25/2$^+$. However, since there is no 
direct experimental evidence for the same, this proposition has not been indicated in the level scheme (Fig. 4) and the 
assignment has not been included in the table (Table I). 
The sharp decrease in the relative intensity of the $\gamma$-ray transitions across the 2156-keV
state is also noteworthy and can be ascribed to the state being an isomer of T$_{1/2}$ $>$ 200 ns \cite{Fan86}. \\

\begin{figure}
\includegraphics[angle=-90,scale=.35,trim=1.0cm 0.0cm 1.0cm 1.0cm,clip=true]{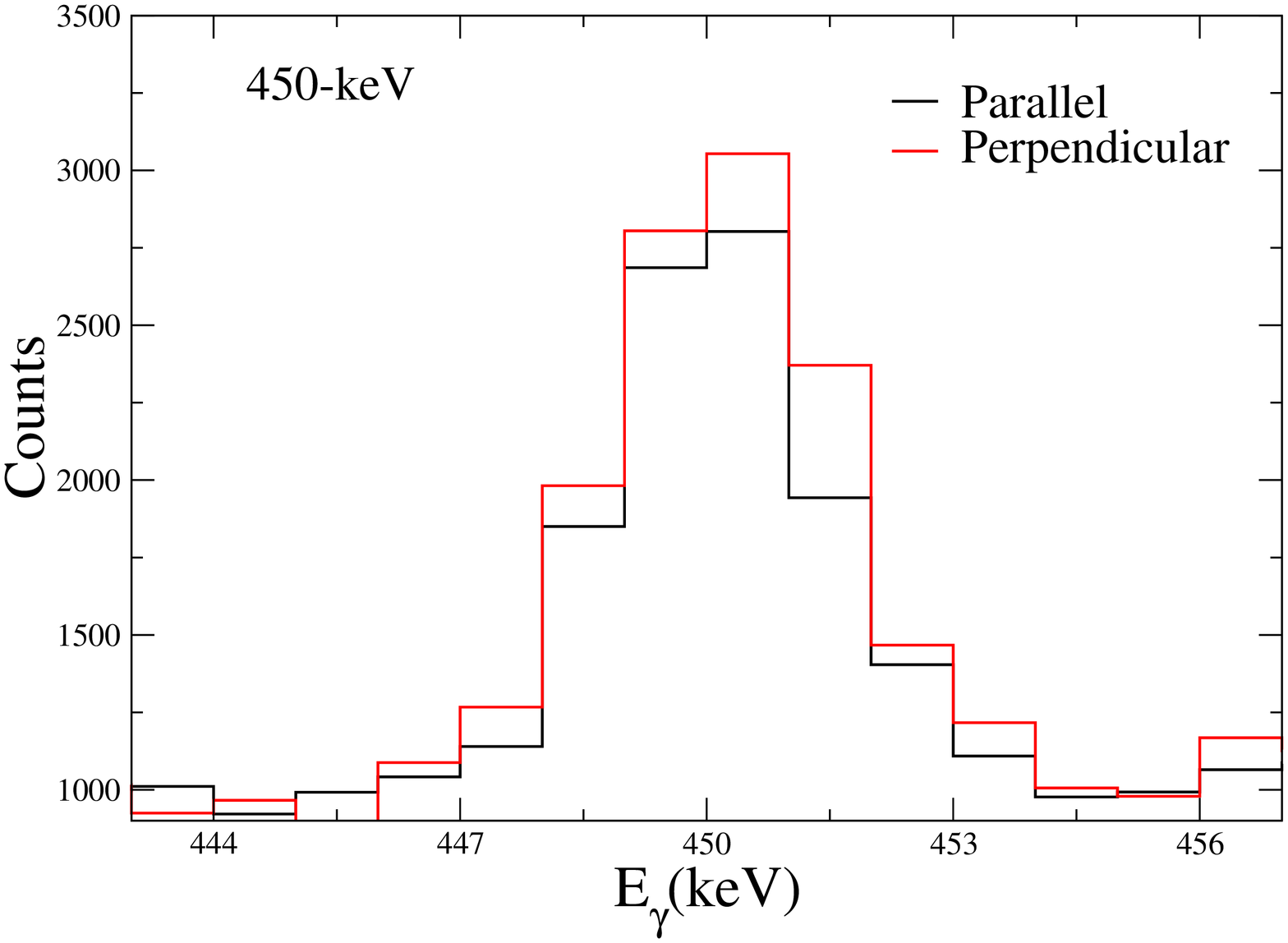}
\caption{\label{fig7}Spectra of 450-keV transition peak corresponding to the perpendicular and the parallel scattering events in the HPGe clover detectors at 90$^o$.}
\end{figure}

Previous studies \cite{Fan86,Fan90} on the Po isotopes had reported a number of isomers therein.
Some of these, with half-lives around few ns, have been reexamined in the current study 
using the centroid shift method \cite{Mad22,Yad22}.
In the present implementation of the technique, the time difference between the feeding and the decaying
$\gamma$-ray transitions of a state is histogrammed alternately by defining one as the start (stop) and the 
other as stop (start). The difference between the centroids of the two distribution is known to be 2$\tau$, 
$\tau$ being the average lifetime of the state. Fig. 8 illustrates the time difference spectra between the
(i) 788- and 262-keV transitions that respectively feeds and de-excites the 3387-keV state in $^{204}$Po \cite{Fan90}, also
populated in the present experiment, and (ii) 637- and 304-keV transitions that respectively feeds and de-excites the
2789-keV state in $^{203}$Po. The half-life of the 3387-keV state in $^{204}$Po was determined 
by Fant {\it{et al.}} \cite{Fan90} as 9$\pm$3 ns, presumably following an analysis of the time profile 
of the decaying transition with respect to the RF of the accelerator. The present analysis has resulted
in T$_{1/2}$ = 6.2$\pm$0.8 ns, that is within the limits of uncertainty on the previous estimate and validates
the present analysis. The latter carried out for the 2789-keV state in $^{203}$Po yields its 
T$_{1/2}$ = 7.1$\pm$0.1 ns that is lesser than the previous value, also reported by Fant {\it{et al.}} \cite{Fan86},
of 12$\pm$2 ns. \\ 

\begin{figure}
\includegraphics[angle=0,scale=.30,trim=1.0cm 1.0cm 0.0cm 1.0cm,clip=true]{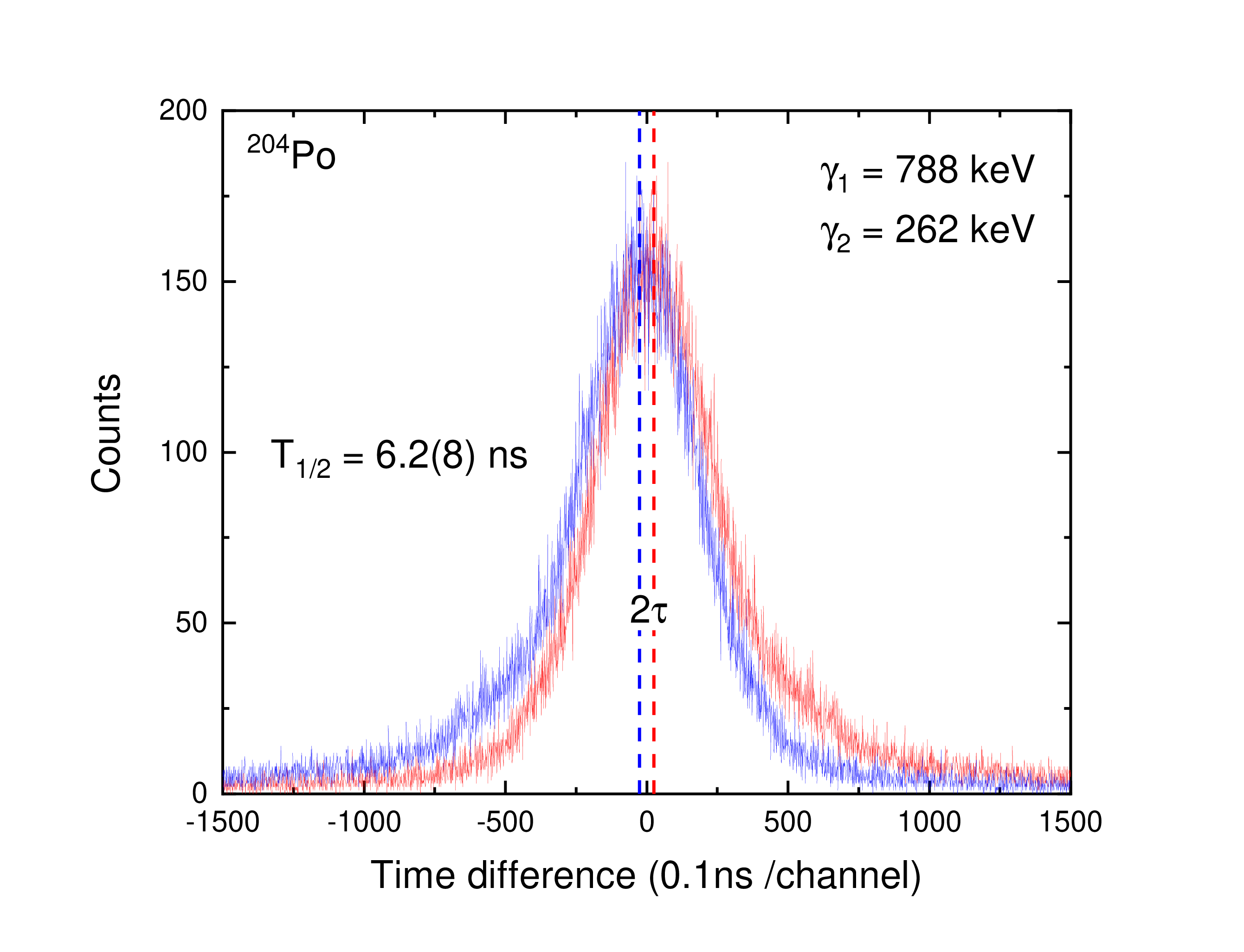}
\includegraphics[angle=0,scale=.30,trim=1.0cm 1.0cm 0.0cm 1.0cm,clip=true]{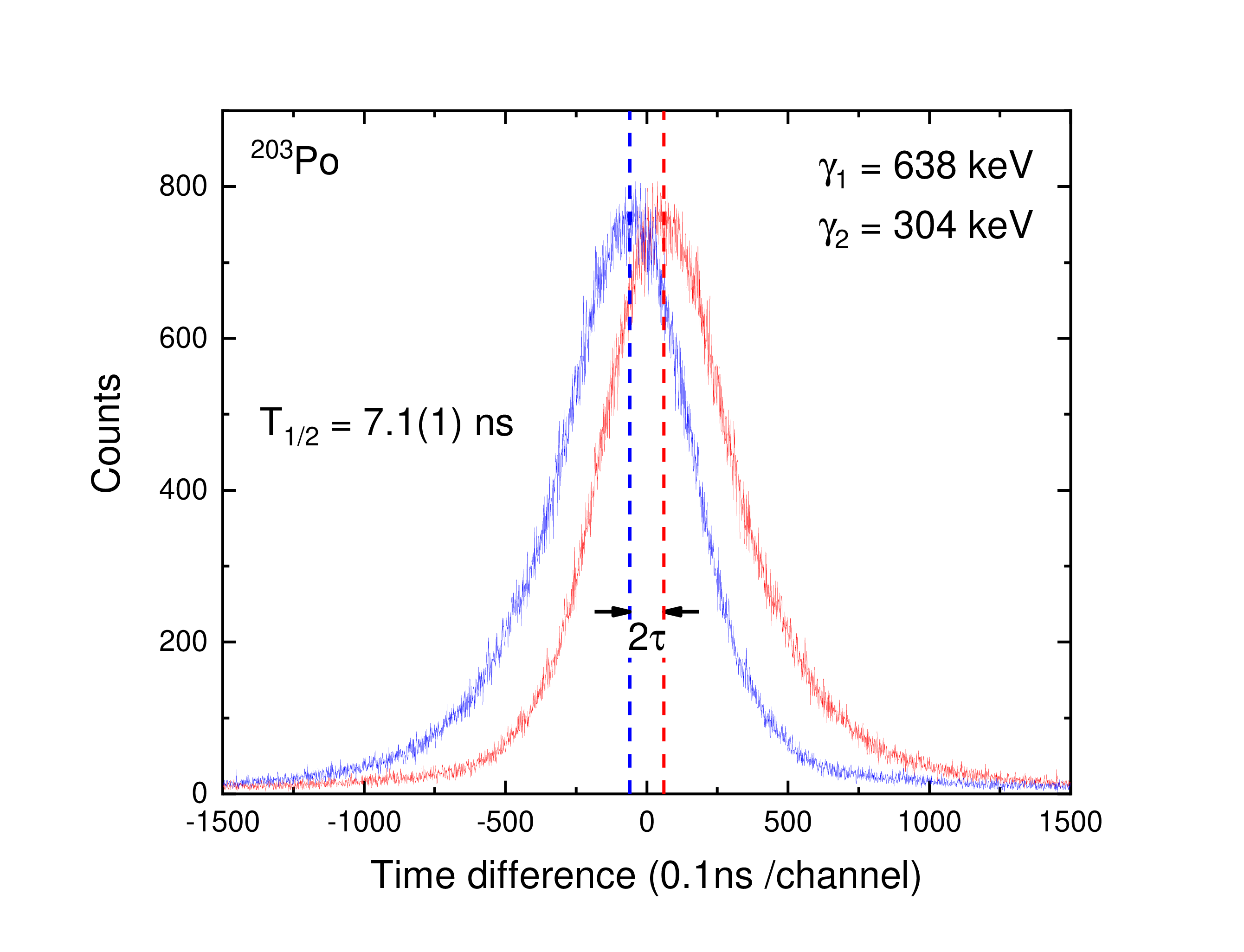}
\caption{\label{fig8}(Color online) Time difference spectra between transitions indicated in the inset, for determining isomeric lifetimes. The upper panel corresponds to the state at 3387-keV state in $^{204}$Po while the bottom panel is for 2789-keV state in $^{203}$Po. (Please refer to the text for details.)}
\end{figure}

The experimentally observed level scheme of the $^{203}$Po nucleus has been interpreted 
through single particle excitations in the framework of the shell model. The same is detailed
in the next section. \\

\LTcapwidth=\textwidth
\begin{longtable*}{ccccccccccc}
\caption{\label{tab1}Details of the levels and the $\gamma$-ray transitions of $^{203}$Po nucleus observed in the present work. The energy of a $\gamma$-ray transition is the weighted average of its value in multiple gates. The relative intensities ($I_\gamma$) of the $\gamma$-ray transitions are normalized with respect to the intensity of 466-keV transition as observed in 612-keV gated spectrum. The ADO ratios($R_{ADO}$), polarization asymmetry ($\Delta_{pol}$), and linear polarization ($P$) of the transitions are determined using the procedure described in Section II. The $N$ and $a$ superscripts indicate the assignments that have been respectively adopted from NNDC \cite{nndc} and/or Fant {\it{et al.}} \cite{Fan86}.} \\
\hline
$E_i (keV) $   & $E_{\gamma} (keV) $    &$I_{\gamma}$ & $J_i^{\pi}$     & $J_f^{\pi}$&$R_{ADO}$  &$\Delta_{pol}$ & P & Multipolarity\\
\hline
\hline
\endfirsthead

\multicolumn{11}{c}%
{{ \tablename \thetable{} -- continued from previous page}} \\
\hline
$E_i (keV) $   & $E_{\gamma} (keV) $   & $I_{\gamma}$ &  $J_i^{\pi}$ & $J_f^{\pi}$&$R_{ADO}$ &$\Delta_{pol}$ & P & Multipolarity\\
\hline
\endhead

\hline
\multicolumn{11}{c}{Continued in next page}\\
\hline
\endfoot
\endlastfoot
638.7  $\pm$0.1    & 577.2$\pm$0.1  &11$\pm$1       & 7/2$^{-}$    & (3/2$^{-})$&               &              & &E2$^N$ \\
                   & 638.7$\pm$0.1  &48$\pm$1       & 7/2$^{-}$    & $5/2^{-}$ &                &              & &M1$^N$ \\
641.7  $\pm$0.2    & 641.7$\pm$0.2  &               & 13/2$^{+}$   & $5/2^{-}$ &                &              & &M4$^N$ \\
1055.2 $\pm$0.1    & 416.5$\pm$0.1  &59$\pm$1       & (11/2$^{-})^a$& $7/2^{-}$&                &              & &M1+E2$^N$ \\
1254.0 $\pm$0.2    & 612.3$\pm$0.1  &               & 17/2$^{+}$   & $13/2^{+}$&                &              & &E2$^N$ \\
1378.8 $\pm$0.3    & 737.1$\pm$0.2  &127$\pm$3      &(17/2$^{+})^a$& $13/2^{+}$&                &              & &(E2)$^N$ \\
1697.1 $\pm$0.8    & 318.3$\pm$0.7  &23$\pm$6       &              &($17/2^{+})^a$&             &              & &     \\
1719.8 $\pm$0.2    & 465.8$\pm$0.1  &1000           & 21/2$^{+}$   & $17/2^{+}$&                &              & &E2$^N$ \\
1974.5 $\pm$0.3    & 595.7$\pm$0.1  &251$\pm$2      &(21/2$^{+})^a$& $(17/2^{+})$&              &              & &E2$^N$ \\
2054.7 $\pm$0.2    & 334.9$\pm$0.1  &643$\pm$14     & 25/2$^{+}$   & $21/2^{+}$  &              &              & &E2$^N$ \\
2077.0 $\pm$0.2    & 356.8$\pm$0.1  &64$\pm$2       & 21/2$^{+}$   & $21/2^{+}$  &              &              & &M1$^N$ \\
2156.4 $\pm$0.3    & 181.9$\pm$0.1  &156$\pm$1&     &$(21/2^{+})^a$&             &              &              &       \\
2184.4 $\pm$0.7    & 805.6$\pm$0.6  &4$\pm$1        &              &($17/2^{+})^a$&             &              & &       \\
2273.6 $\pm$0.2    & 219.1$\pm$0.1  &58$\pm$1       & 27/2$^{+}$   & $25/2^{+}$&0.75$\pm$0.01   &-0.16$\pm$0.09&-0.43$\pm$0.24&M1            \\
2404.2 $\pm$0.2    & 349.2$\pm$0.1  &36$\pm$1       & 25/2$^{+}$   & $25/2^{+}$&                &              & &M1$^N$ \\
                   & 684.1$\pm$0.1  &71$\pm$2       & 25/2$^{+}$   & $21/2^{+}$&                &              & &       \\
2485.8 $\pm$0.2    & 408.6$\pm$0.2  &21$\pm$1       & 23/2$^{+}$   & $21/2^{+}$&                &              & &M1$^N$ \\
                   & 765.9$\pm$0.1  &76$\pm$2       & 23/2$^{+}$   & $21/2^{+}$&                &              & &M1$^N$ \\
2500.2 $\pm$0.2    & 780.1$\pm$0.1  &40$\pm$1       & 23/2$^{+}$   & $21/2^{+}$&                &              & &(M1)$^N$ \\
2523.0 $\pm$0.2    & 468.3$\pm$0.1  &176$\pm$4      & 27/2$^{+}$   & $25/2^{+}$&                &              & &M1$^N$ \\
2527.6 $\pm$0.7    &1148.8$\pm$0.6  &               &              & (17/2$^{+})^a$&            &              & &       \\
2765.1 $\pm$0.4    & 710.4$\pm$0.3  &8$\pm$1        &              & 25/2$^{+}$&                &              & &       \\
2789.1 $\pm$0.2    & 289.3$\pm$0.1  &74$\pm$2       & 25/2$^{-}$   & 23/2$^{+}$&                &              & &(E1)$^N$ \\
                   & 303.5$\pm$0.1  &102$\pm$2      & 25/2$^{-}$   & 23/2$^{+}$&                &              & &E1$^N$  \\
                   & 385.1$\pm$0.1  &54$\pm$2       & 25/2$^{-}$   & 25/2$^{+}$&                &              & &E1$^N$  \\
2820.9 $\pm$0.2    & 297.7$\pm$0.1  &28$\pm$1       & 29/2$^{-}$   & 27/2$^{+}$&                &              & &(E1)$^N$  \\
2863.0 $\pm$0.2    & 589.4$\pm$0.1  &55$\pm$2       & 31/2$^{+}$   & 27/2$^{+}$&1.05$\pm$0.05   &0.05$\pm$0.04 &0.21$\pm$0.17  &E2+M3        \\
2867.2 $\pm$0.2    & 812.5$\pm$0.1  &226$\pm$5      & 29/2$^{+}$   & 25/2$^{+}$&1.34$\pm$0.01   &0.06$\pm$0.01 &0.32$\pm$0.05  &E2           \\
2950.0 $\pm$0.4    & 793.6$\pm$0.2  &               &              &           &0.70$\pm$0.03   &              &               &D            \\
2976.7 $\pm$0.3    & 922.2$\pm$0.2  &13$\pm$1       & 27/2$^{+}$   & 25/2$^{+}$&0.82$\pm$0.06   &-0.05$\pm$0.06&-0.31$\pm$0.37 &M1           \\
3013.5 $\pm$0.2    & 959.1$\pm$0.1  &44$\pm$1       & 27/2$^{+}$   & 25/2$^{+}$&0.83$\pm$0.02   &-0.01$\pm$0.03&-0.06$\pm$0.19 &M1+(E2)      \\
3066.1 $\pm$0.2    & 542.7$\pm$0.1  &46$\pm$1       & 29/2$^{-}$   & 27/2$^{+}$&0.81$\pm$0.01   &0.04$\pm$0.03 &0.16$\pm$0.12  &E1           \\
3093.2 $\pm$0.4    &1038.5$\pm$0.3  &16$\pm$3       & 29/2$^{ }$   & 25/2$^{+}$&1.24$\pm$0.09   &              &               &Q            \\
3107.5 $\pm$0.2    & 584.5$\pm$0.1  &21$\pm$1       & 29/2$^{ }$   & 27/2$^{+}$&0.66$\pm$0.02   &              &               &D            \\
3108.3 $\pm$0.4    & 704.1$\pm$0.3  &11$\pm$1       &              & 25/2$^{+}$&                &              &               &             \\
3220.5 $\pm$0.4    &1165.8$\pm$0.3  &6$\pm$1        &              & 25/2$^{+}$&                &              &               &             \\
3231.8 $\pm$0.2    & 364.5$\pm$0.1  &40$\pm$1       & 31/2$^{+}$   & 29/2$^{+}$&0.79$\pm$0.02   &-0.04$\pm$0.03&-0.13$\pm$0.10 &M1           \\
3236.3 $\pm$0.2    & 368.8$\pm$0.1  &79$\pm$2       & 33/2$^{+}$   & 29/2$^{+}$&1.39$\pm$0.03   &0.03$\pm$0.02 &0.10$\pm$0.06  &E2           \\
3241.9 $\pm$0.2    & 374.7$\pm$0.1  &15$\pm$1       & 33/2$^{+}$   & 29/2$^{+}$&1.41$\pm$0.04   &0.08$\pm$0.10 &0.26$\pm$0.32  &(E2)      \\
3254.4 $\pm$0.2    & 390.9$\pm$0.1  &18$\pm$1       & 33/2$^{-}$   & 31/2$^{+}$&0.97$\pm$0.05   &0.08$\pm$0.07 &0.26$\pm$0.23  &E1+M2        \\
3262.9 $\pm$0.2    & 196.6$\pm$0.1  &16$\pm$1       & 33/2$^{ }$   & 29/2$^{-}$&1.28$\pm$0.04   &              &               &Q            \\
3264.0 $\pm$0.2    & 397.0$\pm$0.1  &22$\pm$1       & 33/2$^{+}$   & 29/2$^{+}$&1.20$\pm$0.04   &0.05$\pm$0.06 &0.17$\pm$0.20  &(E2)      \\
3292.0 $\pm$0.3    &1237.2$\pm$0.2  &5$\pm$1        &              & 25/2$^{+}$&                &              &               &             \\
3416.1 $\pm$0.3    & 627.4$\pm$0.2  &17$\pm$1       &              & 25/2$^{-}$&                &              &               &             \\
3426.5 $\pm$0.2    & 637.3$\pm$0.1  &140$\pm$3      & 29/2$^{-}$   & 25/2$^{-}$&1.39$\pm$0.02   &0.08$\pm$0.01 &0.35$\pm$0.05  &E2           \\
3443.2 $\pm$0.3    & 211.4$\pm$0.2  &29$\pm$1       & 33/2$^{ }$   & 31/2$^{+}$&0.72$\pm$0.02   &              &               &D            \\
3456.2 $\pm$0.2    & 192.3$\pm$0.1  &11$\pm$1       & 35/2$^{ }$   & 33/2$^{+}$&0.68$\pm$0.05   &              &               &D            \\
3488.8 $\pm$0.3    & 234.3$\pm$0.2  &15$\pm$1       & 35/2$^{ }$   & 33/2$^{-}$&0.69$\pm$0.01   &              &               &D         \\
3603.8 $\pm$0.3    &1329.9$\pm$0.2  &9$\pm$1        &              & 27/2$^{+}$&                &              &               &             \\
3711.9 $\pm$0.2    & 844.7$\pm$0.1  &29$\pm$1       & 33/2$^{+}$   & 29/2$^{+}$&1.25$\pm$0.05   &0.04$\pm$0.02 &0.22$\pm$0.11  &E2           \\
3741.7 $\pm$0.2    & 315.1$\pm$0.1  &34$\pm$1       & 31/2$^{-}$   & 29/2$^{-}$&0.98$\pm$0.04   &-0.08$\pm$0.05&-0.24$\pm$0.15 &M1+E2        \\
3876.6 $\pm$0.2    & 450.3$\pm$0.1  &97$\pm$2       & 33/2$^{-}$   & 29/2$^{-}$&1.33$\pm$0.01   &0.07$\pm$0.02 &0.25$\pm$0.07  &E2           \\
3934.7 $\pm$0.3    & 491.5$\pm$0.2  &21$\pm$1       & 37/2$^{ }$   & 33/2$^{ }$&1.36$\pm$0.07   &              &               &Q            \\
3979.4 $\pm$0.3    & 742.9$\pm$0.2  &23$\pm$1       & 35/2$^{+}$   & 33/2$^{+}$&0.77$\pm$0.03   &-0.04$\pm$0.03&-0.20$\pm$0.15 &M1           \\
4146.2 $\pm$0.2    & 891.8$\pm$0.1  &24$\pm$1       & 35/2$^{+}$   & 33/2$^{-}$&0.82$\pm$0.03   &0.01$\pm$0.03 &0.06$\pm$0.18  &(E1)         \\
4352.3 $\pm$0.2    & 475.9$\pm$0.1  &27$\pm$1       & 35/2$^{-}$   & 33/2$^{-}$&0.77$\pm$0.04   &-0.06$\pm$0.02&-0.22$\pm$0.07 &M1           \\
4528.2 $\pm$0.4    & 175.9$\pm$0.3  &13$\pm$1       &              & 35/2$^{-}$&                &              &               &             \\
4612.0 $\pm$0.3    & 735.4$\pm$0.2  &26$\pm$1       & 35/2$^{-}$   & 33/2$^{-}$&0.90$\pm$0.04   &-0.05$\pm$0.07&-0.25$\pm$0.35 &M1+E2        \\
4623.3 $\pm$0.4    & 746.7$\pm$0.3  &8$\pm$1        &              & 33/2$^{-}$&                &              &               &             \\
4645.6 $\pm$0.2    &1189.3$\pm$0.1  &24$\pm$1       & 37/2$^{ }$   & 35/2$^{ }$&0.73$\pm$0.04   &-0.03$\pm$0.07&-0.24$\pm$0.56 &(M1)         \\
4929.8 $\pm$0.5    &1053.2$\pm$0.4  &               &              & 33/2$^{-}$&                &               &               &            \\
\hline
\bigskip
\end{longtable*}

\section{Discussions}

One of the objectives of this endeavor has been to probe the efficacy of the shell model 
in interpreting the excitation scheme of the nuclei in the A $\sim$ 200 region.
There have been similar efforts, in recent times, wherein level structures
of nuclei around the $^{208}$Pb-core are calculated in the shell model framework.
Bothe {\it{et al.}} \cite{Bot22} have reported such calculations for the isomeric states 
in $^{203}$Tl ($Z = 81, N = 122$) while Yadav {\it{et al.}} \cite{Yad22} and 
Madhu {\it{et al.}} \cite{Mad22} have used them for deciphering the particle excitations 
associated with the observed states of $^{215,216}$Fr ($Z = 87, N = 128,129$) nuclei. 
These studies have identified a general overlap, between the experimental and the calculated 
level energies, of within $\sim$ 250-keV as reasonable. \\

Large basis shell model calculation has been carried out in the present work using KHH7B \cite{Her72} Hamiltonian
in the model space spanning $Z = 58-114$ and $N = 100-164$. The latter includes proton orbitals 
$d_{5/2}$, $h_{11/2}$, $d_{3/2}$ and $s_{1/2}$ below $Z = 82$ and the $h_{9/2}$, $f_{7/2}$, and $i_{13/2}$
above; the neutron orbitals are $i_{13/2}$, $p_{3/2}$, $f_{5/2}$, and $p_{1/2}$ below $N = 126$ 
and the $g_{9/2}$, $i_{11/2}$, and $j_{15/2}$ above. Proton excitations across $Z = 82$ closure and neutron excitations
across the closure at $N = 126$ have not been allowed in the calculations. The matrix 
diagonalization has been carried out using the OXBASH \cite{Bro04} code. The comparison between
the calculated and the experimental level energies is illustrated in Figs. 9 and 10. The dominant 
particle configurations along with the energy values of the states are recorded in Table II. \\

\begin{figure}
\includegraphics[angle=-90,scale=.40,trim=0.0cm 0.0cm 0.0cm 0.0cm,clip=true]{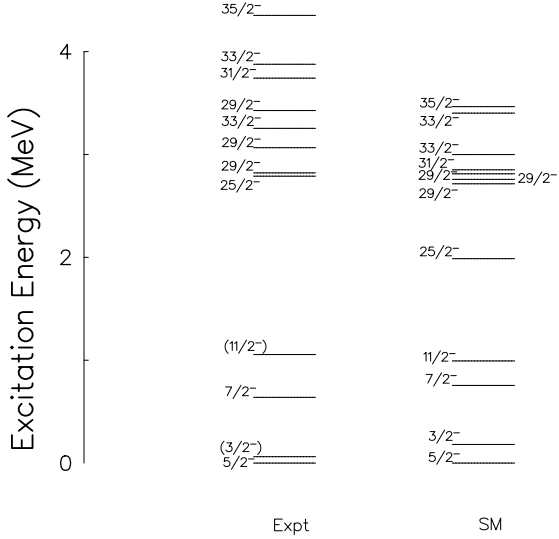}
\caption{\label{fig9}Comparison between the calculated and the experimental level energies of the negative parity states in $^{203}$Po.}
\end{figure}

\begin{figure*}
\includegraphics[angle=-90,scale=.70,trim=0.0cm 0.0cm 5.0cm 0.0cm,clip=true]{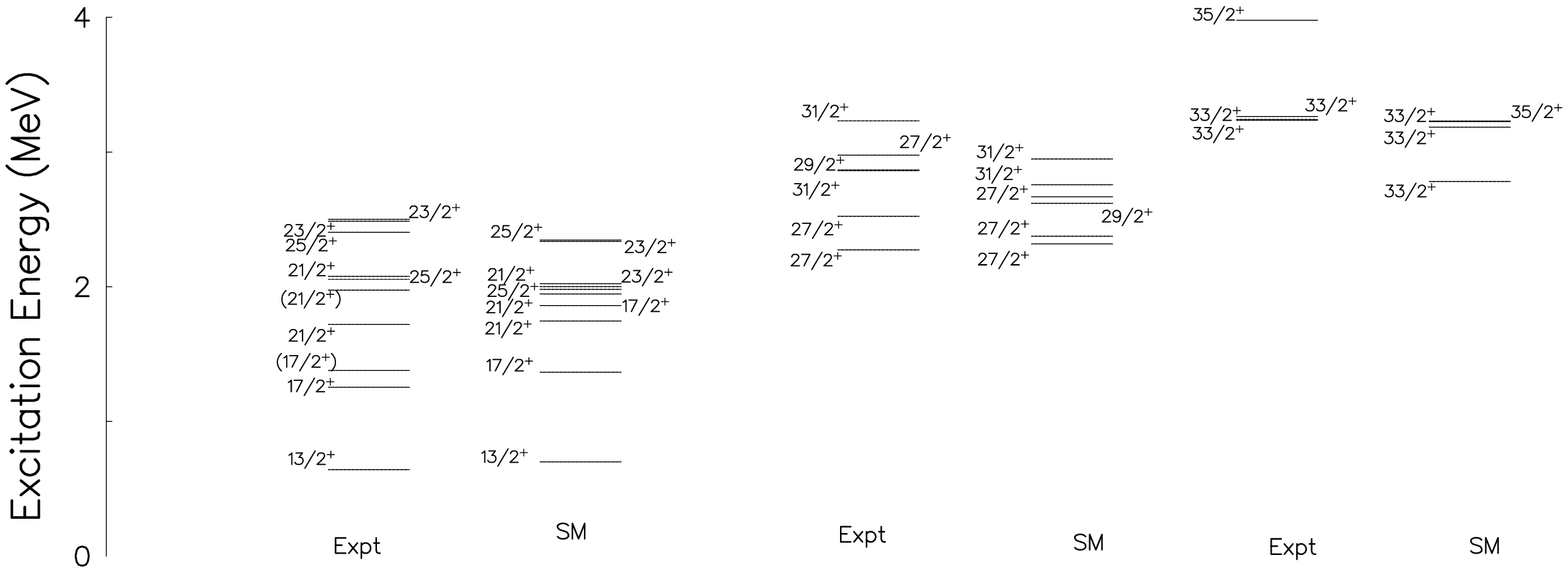}
\caption{\label{fig10}Comparison between the calculated and the experimental level energies of the positive parity states in $^{203}$Po.}
\end{figure*}

The calculated energies of the negative parity states with spin $<$ 29/2 are excellent
overlap with their experimental values, even within $\sim$ 100-keV for some of them.
The 25/2$^-$ level is an exception for which the theoretical and the measured level energies
differ by $\sim$ 800 keV. The dominant particle configurations associated with these states
negative parity states
have been calculated to be $\pi(h_{9/2}^2)\otimes\nu(f_{5/2}^{3,2}p_{3/2}^{2,3}i_{13/2}^{14})$. 
The negative parity states at higher spins, $\ge$ 31/2, are 
poorly represented in the calculations wherein their energies are deviant by as much as
500-keV - 1-MeV with respect to the experimental values. The energy of the calculated yrast 
33/2$^-$ state, however, reasonably overlaps with the measured energy within $\sim$ 250-keV.  
The most probable particle configurations for the negative parity states at higher spins 
correspond to $\pi(h_{9/2}^{1}i_{13/2}^{1})\otimes\nu(f_{5/2}^{2}p_{3/2}^{4}i_{13/2}^{13})$.
However, those of the yrare 33/2$^-$ and the 35/2$^-$ are different 
($\pi(h_{9/2}^{2})\otimes\nu(f_{5/2}^{3}p_{3/2}^{4}i_{13/2}^{12})$ but, as indicated by the
widely deviant calculated energies vis-a-vis the experimental ones, these configurations do not appropriately
represent the relevant states, similar to the other high spin levels of odd parity. \\

\begin{longtable*}{cccccccccccccccc}
\caption{\label{tab2}Main partitions of wave functions of the positive and negative parity states in  $^{203}$Po for KHH7B interaction}    \\
\hline
       &&Level Energy&           & $J^{\pi}$   & Probability & Proton                                            & Neutron         \\
       &&            &           &             &             &         &                                                           \\
 & EXPT&&SM          &           &             &             &         &                                                           \\
\hline
\hline
\endfirsthead

\multicolumn{16}{c}%
{{ \tablename \thetable{} -- continued from previous page}} \\
\hline
       &&Level Energy&           & $J^{\pi}$   & Probability &Proton                                            & Neutron          \\
       &&            &           &             &             &         &                                                           \\
 & EXPT&&SM          &           &             &             &         &                                                           \\
\hline
\hline
\endhead

\hline
\multicolumn{16}{c}{Continued in next page}\\
\hline
\endfoot

\endlastfoot
    &    &    &           &       &                                       & &      &                                            \\
    &    &    &           &       &NEGATIVE PARITY                        & &      &                                            \\
    &    &    &           &       &                                       & &      &                                    \\
 & 0    &&  0        &  $5/2^{-}$  &  29.33  &  $h_{9/2}^2f_{7/2}^0i_{13/2}^0$ &  $f_{5/2}^3p_{3/2}^2p_{1/2}^0i_{13/2}^{14}$     \\
    &&    &&        &             &         &                             &                                                      \\
 & 62   &&  181     &  $(3/2^{-})$ &  39.21  &  $h_{9/2}^2f_{7/2}^0i_{13/2}^0$ &  $f_{5/2}^2p_{3/2}^3p_{1/2}^0i_{13/2}^{14}$      \\
    &&    &&        &             &         &                             &                                           \\
 & 639  &&  755     &  $7/2^{-}$  &  19.96  &  $h_{9/2}^2f_{7/2}^0i_{13/2}^0$ &  $f_{5/2}^2p_{3/2}^3p_{1/2}^0i_{13/2}^{14}$      \\
    &&    &&        &             &         &                             &                                           \\
 & 1055 &&  993     &  $(11/2^{-})$& 37.64  &  $h_{9/2}^2f_{7/2}^0i_{13/2}^0$ &  $f_{5/2}^2p_{3/2}^3p_{1/2}^0i_{13/2}^{14}$     \\
    &&    &&        &             &         &                             &                                           \\
 & 2789 &&  1987    & $25/2^{-}$  &  39.78  &  $h_{9/2}^2f_{7/2}^0i_{13/2}^0$ &  $f_{5/2}^3p_{3/2}^2p_{1/2}^0i_{13/2}^{14}$      \\
    &&    &&        &             &         &                             &                                          \\
 & 2821 &&  2715     & $29/2^{-}$ & 70.95  &  $h_{9/2}^2f_{7/2}^0i_{13/2}^0$ &  $f_{5/2}^3p_{3/2}^2p_{1/2}^0i_{13/2}^{14}$      \\
    &&    &&        &             &         &                             &                                           \\
 & 3066 &&  2758     &  $29/2^{-}$ & 31.65  &  $h_{9/2}^2f_{7/2}^0i_{13/2}^0$ &  $f_{5/2}^3p_{3/2}^4p_{1/2}^0i_{13/2}^{12}$   \\
    &&    &&        &             &         &                             &                                          \\
 & 3254 &&  2999     &  $33/2^{-}$ & 26.06  &  $h_{9/2}^1f_{7/2}^0i_{13/2}^1$ &  $f_{5/2}^2p_{3/2}^4p_{1/2}^0i_{13/2}^{13}$       \\
    &&    &&        &             &         &                             &                                             \\
 & 3427 &&  2812    &  $29/2^{-}$ & 24.93  &  $h_{9/2}^1f_{7/2}^0i_{13/2}^1$ &  $f_{5/2}^2p_{3/2}^4p_{1/2}^0i_{13/2}^{13}$       \\
    &&    &&        &             &         &                             &                                             \\
 & 3742 &&  2851     &  $31/2^{-}$ & 25.53  &  $h_{9/2}^1f_{7/2}^0i_{13/2}^1$ &  $f_{5/2}^2p_{3/2}^4p_{1/2}^0i_{13/2}^{13}$        \\
    &&    &&        &             &         &                             &                                             \\
 & 3877 &&  3402     &  $33/2^{-}$ & 42.34  &  $h_{9/2}^2f_{7/2}^0i_{13/2}^0$ &  $f_{5/2}^3p_{3/2}^4p_{1/2}^0i_{13/2}^{12}$        \\
    &&    &&        &             &         &                             &                                             \\
 & 4352 &&  3465     &  $35/2^{-}$ & 24.51  &  $h_{9/2}^1f_{7/2}^0i_{13/2}^1$ &  $f_{5/2}^2p_{3/2}^4p_{1/2}^0i_{13/2}^{13}$        \\
    &&    &&        &             &         &                             &                                           \\
 & 4612 &&  3671     &  $35/2^{-}$ & 29.32  &  $h_{9/2}^2f_{7/2}^0i_{13/2}^0$ &  $f_{5/2}^3p_{3/2}^4p_{1/2}^0i_{13/2}^{12}$      \\
    &&    &&        &             &         &                             &                                             \\
    &    &    &           &       &                                       &                                             \\
    &    &    &           &       &POSITIVE PARITY                        &                                             \\
    &    &    &           &       &                                       &                                              \\
 & 642  &&  700      &  $13/2^{+}$  &  24.13  &  $h_{9/2}^2f_{7/2}^0i_{13/2}^0$ &  $f_{5/2}^2p_{3/2}^4p_{1/2}^0i_{13/2}^{13}$        \\
    &&    &&            &&        &                                       &                                               \\
 & 1254 &&  1365     &  $17/2^{+}$ &  28.43  &  $h_{9/2}^2f_{7/2}^0i_{13/2}^0$ &  $f_{5/2}^2p_{3/2}^4p_{1/2}^0i_{13/2}^{13}$          \\
    &&    &&           &&         &                                       &                                          \\
 & 1379 &&  1861     &  ($17/2^{+}$) &  23.95  &  $h_{9/2}^2f_{7/2}^0i_{13/2}^0$ &  $f_{5/2}^2p_{3/2}^4p_{1/2}^0i_{13/2}^{13}$          \\
    &&    &&            &&        &                                       &                                          \\
 & 1720 &&  1745     &  $21/2^{+}$ &  23.00  &  $h_{9/2}^2f_{7/2}^0i_{13/2}^0$ &  $f_{5/2}^2p_{3/2}^4p_{1/2}^0i_{13/2}^{13}$     \\
    &&    &&           &&         &                                       &                                          \\
 & 1975 &&  1947     &  ($21/2^{+}$) &30.87  &  $h_{9/2}^2f_{7/2}^0i_{13/2}^0$ &  $f_{5/2}^4p_{3/2}^2p_{1/2}^0i_{13/2}^{13}$     \\
    &&    &&           &&         &                                       &                                          \\
 & 2055 &&  1981     &  $25/2^{+}$ &  32.32  &  $h_{9/2}^2f_{7/2}^0i_{13/2}^0$ &  $f_{5/2}^2p_{3/2}^4p_{1/2}^0i_{13/2}^{13}$   \\
    &&    &&           &&         &                                       &                                          \\
 & 2077 &&  2023     &  $21/2^{+}$ &  27.80  &  $h_{9/2}^2f_{7/2}^0i_{13/2}^0$ &  $f_{5/2}^2p_{3/2}^4p_{1/2}^0i_{13/2}^{13}$   \\
    &&    &&           &&         &                                       &                                          \\
  & 2404 && 2347     &  $25/2^{+}$ &  35.75  &  $h_{9/2}^1f_{7/2}^0i_{13/2}^1$ &  $f_{5/2}^3p_{3/2}^2p_{1/2}^0i_{13/2}^{14}$   \\
    &&    &&           &&         &                                       &                                          \\
 & 2486 &&  2001     &  $23/2^{+}$ &  12.82  &  $h_{9/2}^0f_{7/2}^0i_{13/2}^0$ &  $f_{5/2}^0p_{3/2}^0p_{1/2}^0i_{13/2}^{0}$    \\
    &&    &&           &&         &                                       &                                                  \\
 & 2500 &&  2337     &  $23/2^{+}$ &  19.05  &  $h_{9/2}^0f_{7/2}^0i_{13/2}^0$ &  $f_{5/2}^0p_{3/2}^0p_{1/2}^0i_{13/2}^{0}$    \\
    &&    &&           &&         &                                       &                                          \\
 & 2274 &&  2318     &  $27/2^{+}$ &  22.42  &  $h_{9/2}^1f_{7/2}^0i_{13/2}^1$ &  $f_{5/2}^1p_{3/2}^4p_{1/2}^0i_{13/2}^{14}$   \\
    &&    &&           &&         &                                       &                                          \\
 & 2523 &&  2375     &  $27/2^{+}$ &  26.22  &  $h_{9/2}^2f_{7/2}^0i_{13/2}^0$ &  $f_{5/2}^2p_{3/2}^4p_{1/2}^0i_{13/2}^{13}$    \\
    &&    &&           &&         &                                       &                                           \\
 & 2977 &&  2667     &  $27/2^{+}$ &  31.61  &  $h_{9/2}^2f_{7/2}^0i_{13/2}^0$ &  $f_{5/2}^3p_{3/2}^3p_{1/2}^0i_{13/2}^{13}$     \\
    &&    &&        &             &                                       &                                           \\
 & 2867 &&  2620     &  $29/2^{+}$ &  35.78  &  $h_{9/2}^2f_{7/2}^0i_{13/2}^0$ &  $f_{5/2}^2p_{3/2}^4p_{1/2}^0i_{13/2}^{13}$     \\
    &&    &&        &             &                                       &                                           \\
 & 2863 &&  2758     &  $31/2^{+}$ &  53.28  &  $h_{9/2}^2f_{7/2}^0i_{13/2}^0$ &  $f_{5/2}^2p_{3/2}^4p_{1/2}^0i_{13/2}^{13}$     \\
    &&    &&           &&         &                                       &                                           \\
 & 3232 &&  2949     &  $31/2^{+}$ &  32.75  &  $h_{9/2}^2f_{7/2}^0i_{13/2}^0$ &  $f_{5/2}^4p_{3/2}^2p_{1/2}^0i_{13/2}^{13}$     \\
    &&    &&           &&         &                                       &                                           \\
 & 3236 &&  2782     &  $33/2^{+}$ &  52.68  &  $h_{9/2}^2f_{7/2}^0i_{13/2}^0$ &  $f_{5/2}^2p_{3/2}^4p_{1/2}^0i_{13/2}^{13}$     \\
    &&    &&           &&         &                                       &                                           \\
 & 3242 &&  3185     &  $33/2^{+}$ &  57.96  &  $h_{9/2}^2f_{7/2}^0i_{13/2}^0$ &  $f_{5/2}^3p_{3/2}^3p_{1/2}^0i_{13/2}^{13}$     \\
    &&    &&        &             &                                       &                                           \\
 & 3264 &&  3226     &  $33/2^{+}$ &  30.14  &  $h_{9/2}^2f_{7/2}^0i_{13/2}^0$ &  $f_{5/2}^2p_{3/2}^4p_{1/2}^0i_{13/2}^{13}$     \\
    &&    &&        &             &                                       &                                           \\
 & 3979 &&  3230     &  $35/2^{+}$ &  51.41  &  $h_{9/2}^2f_{7/2}^0i_{13/2}^0$ &  $f_{5/2}^2p_{3/2}^4p_{1/2}^0i_{13/2}^{13}$     \\
    &&    &&        &             &                                       &                                           \\
 & 4146 &&  3517     &  $35/2^{+}$ &  54.27  &  $h_{9/2}^2f_{7/2}^0i_{13/2}^0$ &  $f_{5/2}^3p_{3/2}^3p_{1/2}^0i_{13/2}^{13}$     \\

\hline
\hline
\end{longtable*}

The calculated level energies for most of the positive parity states with spin $<$ 27/2 are
in excellent overlap, within or around 100-keV, with their experimental values. 
The yrare 17/2$^+$ state is an exception for which the calculated and the experimental energies 
differ by $\sim$ 500-keV. However, it is noteworthy that the spin-parity assignment of the
1379-keV state as second 17/2$^+$ was by Fant {\it{et al.}} \cite{Fan86} and is tentative. This could not be
confirmed in the present study. If the parity assignment of the state is changed, it would be the yrast 
(and only observed) 17/2$^-$ level with calculated energy of 1214-keV that is in reasonable 
overlap with the experimental value. It is noted that, in such a scenario, the 737-keV (17/2$^-$ $\rightarrow$ 13/2$^+$)
and 596-keV (21/2$^+$ $\rightarrow$ 17/2$^-$) transitions would be M2 ones and, according to the
Weisskopf estimate, would translate into lifetimes of $\sim$ few ns for the states they de-excite.
These lifetimes are much less than the $\gamma$-$\gamma$ coincidence window (200 ns) of the experiment
and, thus, will not impact the observed intensity of the transitions.
The yrast and the yrare 23/2$^+$ state respectively exhibit differences of $\sim$ 500-keV
and $\sim$ 200-keV between their theoretical and measured values. While the latter can still be perceived as 
a reasonable overlap, a deviation of the calculated energy by $\sim$ 500-keV with respect to the
experimental one indicates an aberrant representation of the state in the framework of the shell model 
calculations. It is also noteworthy that the yrast 23/2$^+$ state is calculated to be
of substantially mixed configurations, compared to the other states of the nucleus, and the
numerically dominant partition is only of 13\% probability. As far as the positive parity states of spin
$\gtrsim$ 29/2 are concerned, the overlap of experimental and calculated energies is 
of considerable variance. While they excellently agree for the yrast 31/2$^+$, the yrare and the 
third 33/2$^+$ levels, within $\sim$ 100-keV, the difference is $\sim$ 250-450 keV for the 29/2$^+$ 
and the yrast 31/2$^+$ and 33/2$^+$. It is still higher, $\sim$ 700-keV, for 35/2$^+$.
Such deviations, at the highest excitations observed in the nucleus, can be ascribed to the
limitations of the model calculations in representing the associated multiparticle 
configurations based on the high-j orbitals (that characterize the relevant model space).
Most of the positive parity states have been calculated to be of dominant configuration 
$\pi(h_{9/2}^{2})\otimes\nu(f_{5/2}^{2-4}p_{3/2}^{2-4}i_{13/2}^{13}$. The exceptions 
are the yrare 25/2$^+$ state, for which the calculated dominant configuration is 
$\pi(h_{9/2}^{1}i_{13/2}^{1})\otimes\nu(f_{5/2}^{3}p_{3/2}^{2}i_{13/2}^{14}$, and the
yrast 27/2$^+$ state, for which the most probable configuration is 
$\pi(h_{9/2}^{1}i_{13/2}^{1})\otimes\nu(f_{5/2}^{1}p_{3/2}^{4}i_{13/2}^{14}$. It is
noteworthy that the calculated and the experimental energies of these states agree within $\sim$ 50-keV
that presumably vindicates the interpretation of their underlying excitations. \\

If the 2156-keV state is assigned a spin-parity of 25/2$^+$, as discussed in the previous section, 
following an E2 assignment for the 182-keV transition (that de-excites the level), the level is then 
the yrare 25/2$^+$ and exhibits a reasonable overlap, within $\sim$ 200 keV, 
with the calculated energy (2347-keV). The current yrare 25/2$^+$ at 2404-keV is then 
the third 25/2$^+$ state and its energy is in excellent agreement with the theoretical value 
of 2383-keV. Once again, since there is no direct experimental evidence to corroborate the
spin-parity assignment of the state at 2156-keV, this has not been included in the table. \\

It may thus be summed up that the observed excitation scheme of the $^{203}$Po nucleus
could be satisfactorily interpreted within the framework of the large basis shell model 
calculations. The specific deviations might have resulted from the limitations
of the Hamiltonian that requires further refinements. The latter is expected to be facilitated
by the availability of experimental data through endeavors such as the present study. \\

\section{Conclusion}

The level structure of the $^{203}$Po nucleus has been probed following its population 
in $^{194}$Pt($^{13}$C,4n) reaction at E$_{lab}$ = 74 MeV. The excitation scheme of the
nucleus has been established upto $\sim$ 5 MeV and spin $\sim$ 18$\hbar$. Twenty new $\gamma$-ray
transitions have been added in the level scheme of the the nucleus and spin-parity assignments 
have been either made or confirmed for a number of states therein. The observed level scheme 
has been satisfactorily interpreted within the framework of large basis shell model calculations
wherein the excited states of the nucleus have been ascribed to proton excitations in
$h_{9/2}$ and $i_{13/2}$ orbitals outside the $Z = 82$ closure and neutron excitations in 
$f_{5/2}$, $p_{3/2}$ and $i_{13/2}$ orbitals in the $N = 126$ shell.
The overlap between the experimental and the calculated level energies, of $^{203}$Po, 
upholds the credibility of the shell model in catering to a microscopic description
of the excitation scheme even for heavy nuclei in the $A \sim 200$ region and in model space
consisting of high-j orbitals. Further refinements in the model calculations are envisaged
to follow the availability of experimental data. \\

\section*{Acknowledgments}

The authors wish to thank the staff associated with the
Pelletron Facility at IUAC, New Delhi, for their help and support during the experiment.
We record our deepest gratitude for Late Prof. Asimananda Goswami and Mr. Pradipta Kumar Das, of 
the Saha Institute of Nuclear Physics (SINP), Kolkata, for their guidance, help
and active contribution in the target preparation.
Help and support received from Mr. Kausik Basu (UGC-DAE CSR, KC) during the experiment, is appreciated. 
Help of V. Vishnu Jyothi during the experiment is also acknowledged.
P.C.S. acknowledges a research grant from SERB (India), CRG/2019/000556.
B.M. acknowledges the support from Department of Science and Technology, Government of India through
DST/INSPIRE Fellowship (IF200310).
U.G. acknowledges the support from US National Science Foundation through Grant No. PHY1762495.
This work is partially supported by the Department of Science
and Technology, Government of India (No. IR/S2/PF-03/2003-II).

\bibliography{203Po_rr}

\begin{thebibliography}{19}
\expandafter\ifx\csname natexlab\endcsname\relax\def\natexlab#1{#1}\fi
\expandafter\ifx\csname bibnamefont\endcsname\relax
  \def\bibnamefont#1{#1}\fi
\expandafter\ifx\csname bibfnamefont\endcsname\relax
  \def\bibfnamefont#1{#1}\fi
\expandafter\ifx\csname citenamefont\endcsname\relax
  \def\citenamefont#1{#1}\fi
\expandafter\ifx\csname url\endcsname\relax
  \def\url#1{\texttt{#1}}\fi
\expandafter\ifx\csname urlprefix\endcsname\relax\def\urlprefix{URL }\fi
\providecommand{\bibinfo}[2]{#2}
\providecommand{\eprint}[2][]{\url{#2}}

\bibitem[{\citenamefont{Rahkonen et~al.}(1985)\citenamefont{Rahkonen, Fant,
  Herrlander, Honkanen, K{\"a}llberg, and Weckstr{\"o}m}}]{Rah85}
\bibinfo{author}{\bibfnamefont{V.}~\bibnamefont{Rahkonen}},
  \bibinfo{author}{\bibfnamefont{B.}~\bibnamefont{Fant}},
  \bibinfo{author}{\bibfnamefont{C.~J.} \bibnamefont{Herrlander}},
  \bibinfo{author}{\bibfnamefont{K.}~\bibnamefont{Honkanen}},
  \bibinfo{author}{\bibfnamefont{A.}~\bibnamefont{K{\"a}llberg}},
  \bibnamefont{and}
  \bibinfo{author}{\bibfnamefont{T.}~\bibnamefont{Weckstr{\"o}m}},
  \bibinfo{journal}{Nucl. \ Phy. \ A} \textbf{\bibinfo{volume}{441}},
  \bibinfo{pages}{11} (\bibinfo{year}{1985}).

\bibitem[{\citenamefont{Weckstr{\"o}m et~al.}(1985)\citenamefont{Weckstr{\"o}m,
  Fant, L{\"o}nnroth, Rahkonen, K{\"a}llberg, and Herrlander}}]{Wec85}
\bibinfo{author}{\bibfnamefont{T.}~\bibnamefont{Weckstr{\"o}m}},
  \bibinfo{author}{\bibfnamefont{B.}~\bibnamefont{Fant}},
  \bibinfo{author}{\bibfnamefont{T.}~\bibnamefont{L{\"o}nnroth}},
  \bibinfo{author}{\bibfnamefont{V.}~\bibnamefont{Rahkonen}},
  \bibinfo{author}{\bibfnamefont{A.}~\bibnamefont{K{\"a}llberg}},
  \bibnamefont{and} \bibinfo{author}{\bibfnamefont{C.~J.}
  \bibnamefont{Herrlander}}, \bibinfo{journal}{Z. \ Phys. \ A}
  \textbf{\bibinfo{volume}{321}}, \bibinfo{pages}{231} (\bibinfo{year}{1985}).

\bibitem[{\citenamefont{Fant et~al.}(1986)\citenamefont{Fant, Weckstr{\"o}m,
  Rahkonen, Herrlander, and K{\"a}llberg}}]{Fan86}
\bibinfo{author}{\bibfnamefont{B.}~\bibnamefont{Fant}},
  \bibinfo{author}{\bibfnamefont{T.}~\bibnamefont{Weckstr{\"o}m}},
  \bibinfo{author}{\bibfnamefont{V.}~\bibnamefont{Rahkonen}},
  \bibinfo{author}{\bibfnamefont{C.~J.} \bibnamefont{Herrlander}},
  \bibnamefont{and}
  \bibinfo{author}{\bibfnamefont{A.}~\bibnamefont{K{\"a}llberg}},
  \bibinfo{journal}{Nucl. \ Phy. \ A} \textbf{\bibinfo{volume}{453}},
  \bibinfo{pages}{77} (\bibinfo{year}{1986}).

\bibitem[{\citenamefont{Muralithar et~al.}(2010)\citenamefont{Muralithar, Rani,
  Kumar, Singh, Das, Gehlot, Golda, Jhingan, Madhavan, Nath et~al.}}]{Mur10}
\bibinfo{author}{\bibfnamefont{S.}~\bibnamefont{Muralithar}},
  \bibinfo{author}{\bibfnamefont{K.}~\bibnamefont{Rani}},
  \bibinfo{author}{\bibfnamefont{R.}~\bibnamefont{Kumar}},
  \bibinfo{author}{\bibfnamefont{R.~P.} \bibnamefont{Singh}},
  \bibinfo{author}{\bibfnamefont{J.~J.} \bibnamefont{Das}},
  \bibinfo{author}{\bibfnamefont{J.}~\bibnamefont{Gehlot}},
  \bibinfo{author}{\bibfnamefont{K.~S.} \bibnamefont{Golda}},
  \bibinfo{author}{\bibfnamefont{A.}~\bibnamefont{Jhingan}},
  \bibinfo{author}{\bibfnamefont{N.}~\bibnamefont{Madhavan}},
  \bibinfo{author}{\bibfnamefont{S.}~\bibnamefont{Nath}}, \bibnamefont{et~al.},
  \bibinfo{journal}{Nucl. \ Instr. \ Meth. \ Phys. \ Res. \ A}
  \textbf{\bibinfo{volume}{622}}, \bibinfo{pages}{281} (\bibinfo{year}{2010}).

\bibitem[{\citenamefont{Das et~al.}(2017)\citenamefont{Das, Samanta,
  Chatterjee, Ghosh, Bhattacharjee, Raut, Ghugre, and Sinha}}]{Das17_2}
\bibinfo{author}{\bibfnamefont{S.}~\bibnamefont{Das}},
  \bibinfo{author}{\bibfnamefont{S.}~\bibnamefont{Samanta}},
  \bibinfo{author}{\bibfnamefont{S.}~\bibnamefont{Chatterjee}},
  \bibinfo{author}{\bibfnamefont{A.}~\bibnamefont{Ghosh}},
  \bibinfo{author}{\bibfnamefont{R.}~\bibnamefont{Bhattacharjee}},
  \bibinfo{author}{\bibfnamefont{R.}~\bibnamefont{Raut}},
  \bibinfo{author}{\bibfnamefont{S.~S.} \bibnamefont{Ghugre}},
  \bibnamefont{and} \bibinfo{author}{\bibfnamefont{A.~K.} \bibnamefont{Sinha}},
  \bibinfo{journal}{Proc. DAE Symp. Nucl. Phys.} \textbf{\bibinfo{volume}{62}},
  \bibinfo{pages}{1066} (\bibinfo{year}{2017}).

\bibitem[{\citenamefont{Bhowmik et~al.}(2001)\citenamefont{Bhowmik, Jain, and
  Biswas}}]{Bho01}
\bibinfo{author}{\bibfnamefont{R.~K.} \bibnamefont{Bhowmik}},
  \bibinfo{author}{\bibfnamefont{A.~K.} \bibnamefont{Jain}}, \bibnamefont{and}
  \bibinfo{author}{\bibfnamefont{D.~C.} \bibnamefont{Biswas}},
  \bibinfo{journal}{Proc. DAE Symp. Nucl. Phys.}
  \textbf{\bibinfo{volume}{44B}}, \bibinfo{pages}{422} (\bibinfo{year}{2001}).

\bibitem[{\citenamefont{Radford}(1995)}]{Rad95}
\bibinfo{author}{\bibfnamefont{D.~C.} \bibnamefont{Radford}},
  \bibinfo{journal}{Nucl. \ Instr. \ Meth. \ Phys. \ Res. \ A}
  \textbf{\bibinfo{volume}{361}}, \bibinfo{pages}{297} (\bibinfo{year}{1995}).

\bibitem[{\citenamefont{Samanta et~al.}(2018)\citenamefont{Samanta, Das,
  Bhattacharjee, Chatterjee, Ghugre, Sinha, Garg, Neelam, Kumar, Jones
  et~al.}}]{Sam18}
\bibinfo{author}{\bibfnamefont{S.}~\bibnamefont{Samanta}},
  \bibinfo{author}{\bibfnamefont{S.}~\bibnamefont{Das}},
  \bibinfo{author}{\bibfnamefont{R.}~\bibnamefont{Bhattacharjee}},
  \bibinfo{author}{\bibfnamefont{S.}~\bibnamefont{Chatterjee}},
  \bibinfo{author}{\bibfnamefont{S.~S.} \bibnamefont{Ghugre}},
  \bibinfo{author}{\bibfnamefont{A.~K.} \bibnamefont{Sinha}},
  \bibinfo{author}{\bibfnamefont{U.}~\bibnamefont{Garg}},
  \bibinfo{author}{\bibnamefont{Neelam}},
  \bibinfo{author}{\bibfnamefont{N.}~\bibnamefont{Kumar}},
  \bibinfo{author}{\bibfnamefont{P.}~\bibnamefont{Jones}},
  \bibnamefont{et~al.}, \bibinfo{journal}{Phys. \ Rev.\ C}
  \textbf{\bibinfo{volume}{97}}, \bibinfo{pages}{014319}
  (\bibinfo{year}{2018}).

\bibitem[{\citenamefont{Samanta et~al.}(2019)\citenamefont{Samanta, Das,
  Bhattacharjee, Chatterjee, Ghugre, Sinha, Garg, Neelam, Kumar, Jones
  et~al.}}]{Sam19}
\bibinfo{author}{\bibfnamefont{S.}~\bibnamefont{Samanta}},
  \bibinfo{author}{\bibfnamefont{S.}~\bibnamefont{Das}},
  \bibinfo{author}{\bibfnamefont{R.}~\bibnamefont{Bhattacharjee}},
  \bibinfo{author}{\bibfnamefont{S.}~\bibnamefont{Chatterjee}},
  \bibinfo{author}{\bibfnamefont{S.~S.} \bibnamefont{Ghugre}},
  \bibinfo{author}{\bibfnamefont{A.~K.} \bibnamefont{Sinha}},
  \bibinfo{author}{\bibfnamefont{U.}~\bibnamefont{Garg}},
  \bibinfo{author}{\bibnamefont{Neelam}},
  \bibinfo{author}{\bibfnamefont{N.}~\bibnamefont{Kumar}},
  \bibinfo{author}{\bibfnamefont{P.}~\bibnamefont{Jones}},
  \bibnamefont{et~al.}, \bibinfo{journal}{Phys. \ Rev.\ C}
  \textbf{\bibinfo{volume}{99}}, \bibinfo{pages}{014315}
  (\bibinfo{year}{2019}).

\bibitem[{\citenamefont{Palit et~al.}(2000)\citenamefont{Palit, Jain, Joshi,
  Nagaraj, Rao, Chintalapudi, and Ghugre}}]{Pal00}
\bibinfo{author}{\bibfnamefont{R.}~\bibnamefont{Palit}},
  \bibinfo{author}{\bibfnamefont{H.~C.} \bibnamefont{Jain}},
  \bibinfo{author}{\bibfnamefont{P.~K.} \bibnamefont{Joshi}},
  \bibinfo{author}{\bibfnamefont{S.}~\bibnamefont{Nagaraj}},
  \bibinfo{author}{\bibfnamefont{B.~V.~T.} \bibnamefont{Rao}},
  \bibinfo{author}{\bibfnamefont{S.~N.} \bibnamefont{Chintalapudi}},
  \bibnamefont{and} \bibinfo{author}{\bibfnamefont{S.~S.}
  \bibnamefont{Ghugre}}, \bibinfo{journal}{Pramana}
  \textbf{\bibinfo{volume}{54}}, \bibinfo{pages}{347} (\bibinfo{year}{2000}).

\bibitem[{nnd()}]{nndc}
\urlprefix\url{www.nndc.bnl.gov/ensdf/}.

\bibitem[{\citenamefont{Semmes et~al.}(1987)\citenamefont{Semmes, Braga, Fink,
  Wood, and Cole}}]{Sem87}
\bibinfo{author}{\bibfnamefont{P.~B.} \bibnamefont{Semmes}},
  \bibinfo{author}{\bibfnamefont{R.~A.} \bibnamefont{Braga}},
  \bibinfo{author}{\bibfnamefont{R.~W.} \bibnamefont{Fink}},
  \bibinfo{author}{\bibfnamefont{J.~L.} \bibnamefont{Wood}}, \bibnamefont{and}
  \bibinfo{author}{\bibfnamefont{J.~D.} \bibnamefont{Cole}},
  \bibinfo{journal}{Nucl. \ Phy. \ A} \textbf{\bibinfo{volume}{464}},
  \bibinfo{pages}{381} (\bibinfo{year}{1987}).

\bibitem[{\citenamefont{Kib\'{e}di et~al.}(2008)\citenamefont{Kib\'{e}di,
  Burrows, Trzhaskovskaya, Davidson, and C.W.~Nestor}}]{Kib08}
\bibinfo{author}{\bibfnamefont{T.}~\bibnamefont{Kib\'{e}di}},
  \bibinfo{author}{\bibfnamefont{T.~W.} \bibnamefont{Burrows}},
  \bibinfo{author}{\bibfnamefont{M.~B.} \bibnamefont{Trzhaskovskaya}},
  \bibinfo{author}{\bibfnamefont{P.~M.} \bibnamefont{Davidson}},
  \bibnamefont{and}
  \bibinfo{author}{\bibfnamefont{J.}~\bibnamefont{C.W.~Nestor}},
  \bibinfo{journal}{Nucl. \ Instr. \ Meth. \ Phys. \ Res. \ A}
  \textbf{\bibinfo{volume}{589}}, \bibinfo{pages}{202} (\bibinfo{year}{2008}).

\bibitem[{\citenamefont{Fant et~al.}(1990)\citenamefont{Fant, Weckstr{\"o}m,
  and K{\"a}llberg}}]{Fan90}
\bibinfo{author}{\bibfnamefont{B.}~\bibnamefont{Fant}},
  \bibinfo{author}{\bibfnamefont{T.}~\bibnamefont{Weckstr{\"o}m}},
  \bibnamefont{and}
  \bibinfo{author}{\bibfnamefont{A.}~\bibnamefont{K{\"a}llberg}},
  \bibinfo{journal}{Phys. \ Scr.} \textbf{\bibinfo{volume}{41}},
  \bibinfo{pages}{652} (\bibinfo{year}{1990}).

\bibitem[{\citenamefont{Madhu et~al.}(2022)\citenamefont{Madhu, Yadav, Deo,
  Pragati, Srivastava, Tandel, Wahid, Kumar, Muralithar, Singh et~al.}}]{Mad22}
\bibinfo{author}{\bibnamefont{Madhu}},
  \bibinfo{author}{\bibfnamefont{K.}~\bibnamefont{Yadav}},
  \bibinfo{author}{\bibfnamefont{A.~Y.} \bibnamefont{Deo}},
  \bibinfo{author}{\bibnamefont{Pragati}},
  \bibinfo{author}{\bibfnamefont{P.~C.} \bibnamefont{Srivastava}},
  \bibinfo{author}{\bibfnamefont{S.~K.} \bibnamefont{Tandel}},
  \bibinfo{author}{\bibfnamefont{S.~G.} \bibnamefont{Wahid}},
  \bibinfo{author}{\bibfnamefont{S.}~\bibnamefont{Kumar}},
  \bibinfo{author}{\bibfnamefont{S.}~\bibnamefont{Muralithar}},
  \bibinfo{author}{\bibfnamefont{R.~P.} \bibnamefont{Singh}},
  \bibnamefont{et~al.}, \bibinfo{journal}{Phys. \ Rev.\ C}
  \textbf{\bibinfo{volume}{105}}, \bibinfo{pages}{034307}
  (\bibinfo{year}{2022}).

\bibitem[{\citenamefont{Yadav et~al.}(2022)\citenamefont{Yadav, Deo, Madhu,
  Pragati, Srivastava, Tandel, Wahid, Kumar, Muralithar, Singh et~al.}}]{Yad22}
\bibinfo{author}{\bibfnamefont{K.}~\bibnamefont{Yadav}},
  \bibinfo{author}{\bibfnamefont{A.~Y.} \bibnamefont{Deo}},
  \bibinfo{author}{\bibnamefont{Madhu}},
  \bibinfo{author}{\bibnamefont{Pragati}},
  \bibinfo{author}{\bibfnamefont{P.~C.} \bibnamefont{Srivastava}},
  \bibinfo{author}{\bibfnamefont{S.~K.} \bibnamefont{Tandel}},
  \bibinfo{author}{\bibfnamefont{S.~G.} \bibnamefont{Wahid}},
  \bibinfo{author}{\bibfnamefont{S.}~\bibnamefont{Kumar}},
  \bibinfo{author}{\bibfnamefont{S.}~\bibnamefont{Muralithar}},
  \bibinfo{author}{\bibfnamefont{R.~P.} \bibnamefont{Singh}},
  \bibnamefont{et~al.}, \bibinfo{journal}{Phys. \ Rev.\ C}
  \textbf{\bibinfo{volume}{105}}, \bibinfo{pages}{034307}
  (\bibinfo{year}{2022}).

\bibitem[{\citenamefont{Bothe et~al.}(2022)\citenamefont{Bothe, Tandel, Wahid,
  Srivastava, Bhoy, Chowdhury, Janssens, Kondev, Carpenter, Lauritsen
  et~al.}}]{Bot22}
\bibinfo{author}{\bibfnamefont{V.}~\bibnamefont{Bothe}},
  \bibinfo{author}{\bibfnamefont{S.~K.} \bibnamefont{Tandel}},
  \bibinfo{author}{\bibfnamefont{S.~G.} \bibnamefont{Wahid}},
  \bibinfo{author}{\bibfnamefont{P.~C.} \bibnamefont{Srivastava}},
  \bibinfo{author}{\bibfnamefont{B.}~\bibnamefont{Bhoy}},
  \bibinfo{author}{\bibfnamefont{P.}~\bibnamefont{Chowdhury}},
  \bibinfo{author}{\bibfnamefont{R.~V.~F.} \bibnamefont{Janssens}},
  \bibinfo{author}{\bibfnamefont{F.~G.} \bibnamefont{Kondev}},
  \bibinfo{author}{\bibfnamefont{M.~P.} \bibnamefont{Carpenter}},
  \bibinfo{author}{\bibfnamefont{T.}~\bibnamefont{Lauritsen}},
  \bibnamefont{et~al.}, \bibinfo{journal}{Phys. \ Rev.\ C}
  \textbf{\bibinfo{volume}{105}}, \bibinfo{pages}{044327}
  (\bibinfo{year}{2022}).

\bibitem[{\citenamefont{Herling and Kuo}(1972)}]{Her72}
\bibinfo{author}{\bibfnamefont{G.~H.} \bibnamefont{Herling}} \bibnamefont{and}
  \bibinfo{author}{\bibfnamefont{T.~T.~S.} \bibnamefont{Kuo}},
  \bibinfo{journal}{Nucl. \ Phy. \ A} \textbf{\bibinfo{volume}{181}},
  \bibinfo{pages}{113} (\bibinfo{year}{1972}).

\bibitem[{\citenamefont{Brown}(2004)}]{Bro04}
\bibinfo{author}{\bibfnamefont{B.~A.} \bibnamefont{Brown}},
  \bibinfo{journal}{MSU NSCL Report} p. \bibinfo{pages}{1289}
  (\bibinfo{year}{2004}).

\end{thebibliography}

\end {document}